# Solution of self-consistent equations for the N$^3$LO nuclear energy density functional in spherical symmetry. The program HOSPHE (v1.00)

B.G. Carlsson[a], J. Dobaczewski[a,b], J. Toivanen[a], P. Veselý[a]

[a] *Department of Physics, P.O. Box 35 (YFL), FI-40014 University of Jyväskylä, Finland*
[b] *Institute of Theoretical Physics, University of Warsaw, Hoża 69, PL-00681 Warsaw, Poland*

**Abstract**

We present solution of self-consistent equations for the N$^3$LO nuclear energy density functional. We derive general expressions for the mean fields expressed as differential operators depending on densities and for the densities expressed in terms of derivatives of wave functions. These expressions are then specified to the case of spherical symmetry. We also present the computer program HOSPHE (v1.00), which solves the self-consistent equations by using the expansion of single-particle wave functions on the spherical harmonic oscillator basis.

## PROGRAM SUMMARY

*Manuscript Title:* Solution of self-consistent equations for the N$^3$LO nuclear energy density functional in spherical symmetry.
*Authors:* B.G. Carlsson, J. Dobaczewski, J. Toivanen, and P. Veselý
*Program Title:* HOSPHE (v1.00)
*Journal Reference:*
*Catalogue identifier:*
*Licensing provisions:* none
*Programming language:* FORTRAN-90
*Operating system:* Linux
*RAM:* 50 MB
*Number of processors used:* 1

*External routines/libraries:* LAPACK, BLAS
*Nature of problem:*
The nuclear mean-field methods constitute principal tools of a description of nuclear states in heavy nuclei. Within the Local Density Approximation with gradient corrections up to $N^3LO$ [1], the nuclear mean-field is local and contains derivative operators up to sixth order. The locality allows for an effective and fast solution of the self-consistent equations.
*Solution method:*
The program uses the spherical harmonic oscillator basis to expand single-particle wave functions of neutrons and protons for the nuclear state being described by the $N^3LO$ nuclear energy density functional [1]. The expansion coefficients are determined by the iterative diagonalization of the mean-field Hamiltonian, which depends non-linearly on the local neutron and proton densities.
*Restrictions:*
Solutions are limited to spherical symmetry. The expansion on the harmonic-oscillator basis does not allow for a precise description of asymptotic properties of wave functions.
*Running time:*
50 sec. of CPU time for the ground-state of $^{208}$Pb described by using the $N_0 = 50$ maximum harmonic-oscillator shell included in the basis.

**LONG WRITE-UP**

## 1. Introduction

The nuclear mean-field methods constitute principal tools of a description of nuclear states in heavy nuclei [1]. Their applicability to nuclei is interpreted within the Hohenberg-Kohn [2] and Kohn-Sham theorems [3] involving the nuclear energy density functional (EDF). Recently, we formulated the $\mathrm{N^3LO}$ nuclear EDF with gradient corrections up to sixth order [4]. The present study presents practical formulation of the method, allowing for a solution of the corresponding self-consistent equations. We also present the computer program HOSPHE (v1.00), which solves the self-consistent equations by using the expansion of single-particle wave functions on the spherical harmonic oscillator (HO) basis.

The paper is organized as follows. In Section 2, we present concise review of the method. In Section 3, we give general forms of the $\mathrm{N^3LO}$ potentials, fields, and densities, which are then in Section 4 specified to the case of spherical symmetry. Sections 5–10 describe the structure, installation, and test runs of the code HOSPHE (v1.00), and Section 11 concludes our study.

## 2. Overview of the method

This introductory section is intended as a guide to the subsequent sections, where more detailed derivations and results are presented. Here, we use abbreviated notations so as to give a brief outline of the method, while referring the reader to the following sections for details.

The mean-field eigenvalue equation is obtained by considering the variation of the energy with the condition of the single-particle wavefunctions $\phi_i\left(\boldsymbol{r},\sigma\right)$ to be normalized to unity,

$$\frac{\delta}{\delta\phi_i^*\left(\boldsymbol{r}',\sigma'\right)}\left(E-\sum_i\epsilon_i\int\left|\phi_i\right|^2\mathrm{d}^3\boldsymbol{r}\right)=0. \tag{1}$$

The potential energy is expressed as the EDF of Ref. [4], that is,

$$\mathcal{E}=\int\sum_{a\alpha\beta}C^{\beta}_{a,\alpha}T^{\beta}_{a,\alpha}(\boldsymbol{r})\mathrm{d}^3\boldsymbol{r}, \tag{2}$$

where the grouped indices, such as the greek indices $\alpha = \{n_\alpha L_\alpha v_\alpha J_\alpha\}$ or $\beta = \{n_\beta L_\beta v_\beta J_\beta\}$ and the roman indices $a = \{m_a I_a\}$, denote all the quantum numbers of the local primary $\rho_\beta(\boldsymbol{r})$ and secondary $\rho_{a,\alpha,J}(\boldsymbol{r})$ densities [4]. In Eq. (2), $T^\beta_{a,\alpha}(\boldsymbol{r})$ denotes terms of the functional,

$$T^\beta_{a,\alpha} \equiv [\rho_\beta \rho_{a,\alpha}]_0 \equiv \left[\rho_\beta \left[D_a \rho_\alpha\right]_{J_\beta}\right]_0 \equiv \left[\rho_{n_\beta L_\beta v_\beta J_\beta} \left[D_{m_a I_a} \rho_{n_\alpha L_\alpha v_\alpha J_\alpha}\right]_{J_\beta}\right]_0, \quad (3)$$

$C^\beta_{a,\alpha}$ denote the coupling constants, $D_a$ denote the higher order derivative operators [4], and the sum runs over all terms of the functional. Although not shown explicitly, the sum contains both isoscalar and isovector terms. At present we have neglected neutron-proton mixing, which means that only the $T_z = 0$ component of the isovector densities are present. The convention adopted is such that the isovector contributions to the densities are taken as neutron minus proton densities and isoscalar densities are sums of neutron and proton densities.

For each term of the functional, the variation with respect to the local densities, followed by the integration by parts and recoupling gives,

$$\delta \int T^\beta_{a,\alpha} \mathrm{d}^3\boldsymbol{r} = \int \left( \left[\delta\rho_\beta \left[D_a \rho_\alpha\right]_{J_\beta}\right]_0 + (-1)^{J_\beta - J_\alpha} \left[\delta\rho_\alpha \left[D_a \rho_\beta\right]_{J_\alpha}\right]_0 \right) \mathrm{d}^3\boldsymbol{r}, \quad (4)$$

cf. Eqs. (43) and (44). The primary densities can be expressed in terms of the single-particle wavefunctions as:

$$\rho_{nLvJ} = \left\{ \left[ K_{nL} \sum_{i\sigma\sigma'} \phi_i\left(\boldsymbol{r}, \sigma\right) \sigma_{v;\sigma'\sigma} \phi_i^*\left(\boldsymbol{r}', \sigma'\right) \right]_J \right\}_{\boldsymbol{r}'=\boldsymbol{r}}. \quad (5)$$

The higher order derivative operators $K_{nL}$ [4] are built by coupling the relative-momentum operators $\boldsymbol{k} = \frac{1}{2i}\left(\boldsymbol{\nabla}^\phi - \boldsymbol{\nabla}^{\phi^*}\right)$, where we have used the subscripts to indicate on which function the operators act upon. This allows us to perform the variation with respect to $\phi_i^*$,

$$\begin{aligned} \frac{\delta}{\delta\phi_i^*} \int T^\beta_{a,\alpha} \mathrm{d}^3\boldsymbol{r} &= \left[ \left[K'_{n_\beta L_\beta} \sigma_{v_\beta} \phi_i\right]_{J_\alpha} \left[D_a \rho_\alpha\right]_{J_\alpha} \right]_0 \\ &+ (-1)^{J_\beta - J_\alpha} \left[ \left[K'_{n_\alpha L_\alpha} \sigma_{v_\alpha} \phi_i\right]_{J_\beta} \left[D_a \rho_\beta\right]_{J_\beta} \right]_0. \end{aligned} \quad (6)$$

After performing the variation, the integral above was partially integrated so that the derivatives would not act on the variation of $\phi_i^*$. Therefore, the $K'_{nL}$ operators are built by coupling the relative-momentum operators $\boldsymbol{k}' = \frac{1}{2i}\left(\boldsymbol{\nabla}^{\phi} + \boldsymbol{\nabla}\right)$, where $\boldsymbol{\nabla}$ acts on all the functions of position standing to the right.

The operator on the right-hand-side of Eq. (6) is a formal expression for the mean-field operator. All what remains to be done is to disentangle the gradients $\boldsymbol{\nabla}^{\phi}$ and $\boldsymbol{\nabla}$ from one another – this procedure is performed in Eqs. (55)–(61) below. Finally, the mean-field operator $h\left(\rho\right)$ acquires the form:

$$h\left(\rho\right) = \sum_{\gamma} \left[U_{\gamma}\left[D_{n_{\gamma}L_{\gamma}}\sigma_{v_{\gamma}}\right]_{J_{\gamma}}\right]_{0}, \qquad (7)$$

where the differential operators $D_{n_{\gamma}L_{\gamma}}$ and Pauli matrices $\sigma_{v_{\gamma}}$ act on the single-particle wave functions, and the potentials $U_{\gamma}(\boldsymbol{r})$ are linear combinations of the secondary densities:

$$U_{\gamma} = \sum_{a\alpha\beta;d\delta} C^{\beta}_{a,\alpha}\chi^{\beta;d\delta}_{a,\alpha;\gamma}\left[D_{d}\rho_{\delta}\right]_{J_{\gamma}}. \qquad (8)$$

The coefficients $\chi^{\beta;d\delta}_{a,\alpha;\gamma}$ can be derived by using the recoupling rules presented in Section 3.3. An alternative method, which was also used when building the code HOSPHE (v1.00), was to construct the fields by starting from Eq. (6) and putting them equal to those of Eq. (7). This gives a linear system of equations that can be solved for the unknown coefficients $\chi^{\beta;d\delta}_{a,\alpha;\gamma}$. At $N^3$LO, only 1494 such coefficients are needed, so they can easily be precalculated and stored.

It is now clear, that the key operators in the mean field are given by

$$F_{d\delta,\gamma} = \left[\left[\rho_{d,\delta}\right]_{J_{\gamma}}\left[D_{n_{\gamma}L_{\gamma}}\sigma_{v_{\gamma}}\right]_{J_{\gamma}}\right]_{0}, \qquad (9)$$

and their matrix elements in the single-particle basis $\phi_i\left(\boldsymbol{r},\sigma\right)$ read,

$$\begin{aligned} F^{ii'}_{d\delta,\gamma} &= \langle\phi_i|F_{d\delta,\gamma}|\phi_{i'}\rangle \\ &= \int\sum_{\sigma\sigma'}\phi^{*}_{i}\left(\boldsymbol{r},\sigma\right)\left[\left[\rho_{d,\delta}\right]_{J_{\gamma}}\left[D_{n_{\gamma}L_{\gamma}}\sigma_{v_{\gamma};\sigma\sigma'}\phi_{i'}\left(\boldsymbol{r},\sigma'\right)\right]_{J_{\gamma}}\right]_{0}\mathrm{d}^{3}\boldsymbol{r}. \qquad (10)\end{aligned}$$

Then, the mean-field matrix elements can be written as the following sum:

$$\langle\phi_i|h(\rho)|\phi_{i'}\rangle = \sum_{a\alpha\beta;d\delta\gamma} C^{\beta}_{a,\alpha}\chi^{\beta;d\delta}_{a,\alpha;\gamma}F^{ii'}_{d\delta,\gamma}. \qquad (11)$$

Matrix elements in a spherical basis are derived in Sections 4.4 and 4.5.

When constructing potentials (8), we need expressions to calculate all secondary densities. These can be written as [see Eqs. (76)–(78)]:

$$\rho_{d,\delta,JM} \equiv [D_d\rho_\delta]_{JM} = \sum_{bb'W} A^{bb',W}_{d,\delta,J}\,\rho^{bb',W}_{v_\delta JM}, \tag{12}$$

with

$$\rho^{bb',W}_{v_\delta JM}(\boldsymbol{r}_1) = \left\{\left[\left[D^{(1)}_b D^{(2)}_{b'}\right]_W \rho_{v_\delta}(\boldsymbol{r}_1,\boldsymbol{r}_2)\right]_{JM}\right\}_{\boldsymbol{r}=\boldsymbol{r}_2=\boldsymbol{r}_1}, \tag{13}$$

where the superscripts on the derivative operators indicate on which coordinate they act. The coefficients $A$ can be obtained by explicitly constructing the left- and right-hand sides of Eq. (12), which gives a linear system of equations in derivatives of the density matrix that can be solved for the unknown coefficients $A^{bb',W}_{d,\delta,J}$. At N$^3$LO, only 3138 such coefficients are needed, so they can easily be precalculated and stored. In Section 3.6 we also show how to derive these coefficients by using the recoupling rules and in Section 4.2 we give the expressions for densities in the spherical HO basis.

## 3. General forms of the N$^3$LO potentials, fields, and densities

### *3.1. Building blocks*

We begin by recalling definitions that are used to construct operators and densities in the spin and position space. The basic building blocks are given as in Eqs. (8)–(9) of [4], i.e.,

$$\sigma_{v=0,0} = \sigma_0, \tag{14}$$

$$\sigma_{v=1,\mu=\{-1,0,1\}} = -i\left\{\tfrac{1}{\sqrt{2}}(\sigma_x - i\sigma_y), \sigma_z, \tfrac{-1}{\sqrt{2}}(\sigma_x + i\sigma_y)\right\}, \tag{15}$$

$$\nabla_{1,\mu=\{-1,0,1\}} = -i\left\{\tfrac{1}{\sqrt{2}}(\nabla_x - i\nabla_y), \nabla_z, \tfrac{-1}{\sqrt{2}}(\nabla_x + i\nabla_y)\right\} \tag{16}$$

$$k_{1,\mu=\{-1,0,1\}} = -i\left\{\tfrac{1}{\sqrt{2}}(k_x - ik_y), k_z, \tfrac{-1}{\sqrt{2}}(k_x + ik_y)\right\}, \tag{17}$$

where $\boldsymbol{k}$ is the relative momentum operator:

$$\boldsymbol{k} = \frac{1}{2i}(\boldsymbol{\nabla} - \boldsymbol{\nabla}'). \tag{18}$$

All possible N$^3$LO differential operators $D_{nLM}$, which can be built of gradients (16), are given in the Table I of Ref. [4], where $n$ is the order of the operator and $L$ is its rank with magnetic projection $M$. Exactly in the same way, in Ref. [4] we defined the operators $K_{nLM}$, which are spherical tensors built of the relative momentum operators $k$ (17).

Hermitian-conjugation properties of the building blocks read:

$$\sigma^+_{v\mu} = Q_\sigma(-1)^{v-\mu}\sigma_{v,-\mu} \quad \text{for} \quad Q_\sigma = +1, \tag{19}$$

$$\nabla^+_{1\mu} = Q_\nabla(-1)^{1-\mu}\nabla_{1,-\mu} \quad \text{for} \quad Q_\nabla = -1, \tag{20}$$

$$k^+_{1\mu} = Q_k(-1)^{1-\mu}k_{1,-\mu} \quad \text{for} \quad Q_k = +1. \tag{21}$$

For any pair of *commuting* operators $A_{\lambda\mu}$ and $B_{\lambda\mu}$ that have the following hermitian-conjugation properties:

$$A^+_{\lambda\mu} = Q_A(-1)^{\lambda-\mu}A_{\lambda,-\mu}, \tag{22}$$

$$B^+_{\lambda'\mu'} = Q_B(-1)^{\lambda'-\mu'}B_{\lambda',-\mu'}, \tag{23}$$

$$\tag{24}$$

the operator $C_{LM}$ built by the angular momentum coupling,

$$C_{LM} \equiv [A_\lambda B_{\lambda'}]_{LM} = \sum_{\mu\mu'} C^{LM}_{\lambda\mu\lambda'\mu'} A_{\lambda\mu}B_{\lambda'\mu'}, \tag{25}$$

behaves under the hermitian conjugation as:

$$C^+_{LM} = Q_C(-1)^{L-M}C_{L,-M} \quad \text{for} \quad Q_C = Q_AQ_B. \tag{26}$$

As a consequence, we have

$$D^+_{nLM} = (-1)^n \quad (-1)^{L-M}D_{nL,-M}, \tag{27}$$

$$K^+_{nLM} = (-1)^{L-M}K_{nL,-M}. \tag{28}$$

We note that the gradient operators (16) and (17) obey the Biedenharn-Rose phase conventions of

$$\nabla^*_{1\mu} = (-1)^{1-\mu}\nabla_{1,-\mu}, \tag{29}$$

$$k^*_{1\mu} = - \quad (-1)^{1-\mu}k_{1,-\mu}, \tag{30}$$

which gives

$$D^*_{nLM} = (-1)^{L-M}D_{nL,-M}, \tag{31}$$

$$D^+_{nLM} = (-1)^n D^*_{nLM}, \tag{32}$$

$$D^T_{nLM} = (-1)^n D_{nLM}, \tag{33}$$

and

$$\begin{aligned} K^{*}_{nLM} &= (-1)^{n}(-1)^{L-M}K_{nL,-M}, &(34)\\ K^{+}_{nLM} &= (-1)^{n}K^{*}_{nLM}, &(35)\\ K^{T}_{nLM} &= (-1)^{n}K_{nLM}, &(36)\end{aligned}$$

where superscript $T$ denotes the transposed operator.

*3.2. Potentials*

The potential energy to be varied over the wave functions is given in Ref. [4] and reads

$$\mathcal{E} = \int \mathrm{d}^3\boldsymbol{r}\mathcal{H}(\boldsymbol{r}), \tag{37}$$

for

$$\begin{aligned} \mathcal{H}(\boldsymbol{r}) &= \sum_{\substack{n'L'v'J'\\ mI,nLvJ}} C^{n'L'v'J'}_{mI,nLvJ}\left[\rho_{n'L'v'J'}(\boldsymbol{r})[D_{mI}\rho_{nLvJ}(\boldsymbol{r})]_{J'}\right]_0 \\ &= \sum_{\substack{n'L'v'J'M'\\ mI,nLvJ}} C^{n'L'v'J'}_{mI,nLvJ}\frac{(-1)^{J'-M'}}{\sqrt{2J'+1}}\,\rho_{n'L'v'J'M'}(\boldsymbol{r})[D_{mI}\rho_{nLvJ}(\boldsymbol{r})]_{J',-M'}, \end{aligned} \tag{38}$$

where $C^{n'L'v'J'}_{mI,nLvJ}$ are the coupling constants and $\rho_{nLvJ}(\boldsymbol{r})$ are the primary densities:

$$\rho_{nLvJ}(\boldsymbol{r}) = \{[K_{nL}\rho_v(\boldsymbol{r},\boldsymbol{r}')]_J\}_{\boldsymbol{r}'=\boldsymbol{r}}\,, \tag{39}$$

which are built by acting with the relative momentum operators $K_{nLM}$ on the scalar ($v=0$) and vector ($v=1$) non-local densities:

$$\rho_{v\mu}\left(\boldsymbol{r},\boldsymbol{r}'\right) = \sum_{\sigma\sigma'}\rho\left(\boldsymbol{r}\sigma,\boldsymbol{r}'\sigma'\right)\langle\sigma'\left|\sigma_{v\mu}\right|\sigma\rangle\,. \tag{40}$$

Note that the sum in Eq. (38) runs over the indices ordered in a specific way, defined in Ref. [4, 5], namely,

$$\{n'L'v'J'\} \leq \{nLvJ\}. \tag{41}$$

We first vary $\mathcal{E}$ over the densities and then, in Section 3.3, we vary densities over the wave functions, that is, we begin with

$$\delta\mathcal{E} = \int \mathrm{d}^3\boldsymbol{r}\delta\mathcal{H}(\boldsymbol{r}) = \sum_{n'L'v'J'M'}\int \mathrm{d}^3\boldsymbol{r}\frac{\partial\mathcal{H}}{\partial\rho_{n'L'v'J'M'}}(\boldsymbol{r})\delta\rho_{n'L'v'J'M'}. \tag{42}$$

An explicit variation over the primary densities under the differential operators $D_{mI}$ can be avoided by first integrating by parts and recoupling. The recoupling within a scalar [6] is simple, namely,

$$[A_{J'}[B_I C_J]_{J'}]_0 \;=\; (-1)^{J'+I-J}[C_J[B_I A_{J'}]_J]_0. \tag{43}$$

Hence, the integration by parts gives:

$$\begin{aligned}\mathcal{H}(\boldsymbol{r}) &= \sum_{\substack{n'L'v'J'\\ mI,nLvJ}} (-1)^{J'+I-J} C^{n'L'v'J'}_{mI,nLvJ}\,[\rho_{nLvJ}(\boldsymbol{r})[D^T_{mI}\rho_{n'L'v'J'}(\boldsymbol{r})]_J]_0\\ &= \sum_{\substack{n'L'v'J'\\ mI,nLvJ}} (-1)^{m+J'+I-J} C^{n'L'v'J'}_{mI,nLvJ}\,[\rho_{nLvJ}(\boldsymbol{r})[D_{mI}\rho_{n'L'v'J'}(\boldsymbol{r})]_J]_0\\ &= \sum_{\substack{n'L'v'J'\\ mI,nLvJ}} (-1)^{J-J'} C^{nLvJ}_{mI,n'L'v'J'}\,[\rho_{n'L'v'J'}(\boldsymbol{r})[D_{mI}\rho_{nLvJ}(\boldsymbol{r})]_{J'}]_0, \end{aligned} \tag{44}$$

where we have used Eq. (33), then we changed the names of indices, and we also used the fact that $m+I$ is even.

Therefore, the variation under the differential operators $D_{mI}$ only gives the transposition of indices $\{n'L'v'J'\} \leftrightarrow \{nLvJ\}$ and the phase. The complete variation of the energy then reads:

$$\delta\mathcal{E} \;=\; \sum_{n'L'v'J'} \int \mathrm{d}^3\boldsymbol{r}[\delta\rho_{n'L'v'J'}\tilde{U}_{n'L'v'J'}(\boldsymbol{r})]_0, \tag{45}$$

where we defined potentials

$$\tilde{U}_{n'L'v'J'M'}(\boldsymbol{r}) = \sum_{mI,nLvJ} \left\{ C^{n'L'v'J'}_{mI,nLvJ} + (-1)^{J-J'} C^{nLvJ}_{mI,n'L'v'J'} \right\} [D_{mI}\rho_{nLvJ}(\boldsymbol{r})]_{J'M'}. \tag{46}$$

Note that because of the ordering (41), in Eq. (46) either the first or the second term is non-zero (for $\{n'L'v'J'\} \neq \{nLvJ\}$), or both terms add up to $2C^{n'L'v'J'}_{mI,nLvJ}$ (for $\{n'L'v'J'\} = \{nLvJ\}$).

*3.3. Fields*

Now we are in a position to perform the variation of densities over the wave functions. To this end, we assume that the non-local density matrices in Eq. (40) have the general form of

$$\rho\,(\boldsymbol{r}\sigma, \boldsymbol{r}'\sigma') = \sum_i \phi_i(\boldsymbol{r}\sigma)\psi_i(\boldsymbol{r}'\sigma'). \tag{47}$$

This form allows us to carry out the derivation for several important specific cases simultaneously. Namely, for the standard HF case one has:

$$\psi_i(\boldsymbol{r}'\sigma') = \phi_i^*(\boldsymbol{r}'\sigma'), \tag{48}$$

and the sum runs over the occupied states only, $i = 1, \ldots, A$. Similarly, for the BCS case, or for the HFB case in the canonical basis, one has:

$$\psi_i(\boldsymbol{r}'\sigma') = v_i^2\phi_i^*(\boldsymbol{r}'\sigma'), \tag{49}$$

where $v_i^2$ are the occupation factors and the sum runs over the pairing window. For transition densities pertaining to the symmetry restoration, one has:

$$\psi_i(\boldsymbol{r}'\sigma') = \sum_j O_{ij}^{-1}(\alpha)\phi_j^*(\boldsymbol{r}'\sigma', \alpha), \tag{50}$$

Where $\phi_j(\boldsymbol{r}'\sigma', \alpha) = R(\alpha)\phi_j(\boldsymbol{r}'\sigma')$ are the wave functions transformed by the symmetry operator $R(\alpha)$ and $O_{ij}(\alpha)$ is the overlap matrix:

$$O_{ij}(\alpha) = \int \mathrm{d}^3\boldsymbol{r} \sum_\sigma \phi_i^*(\boldsymbol{r}\sigma, \alpha)\phi_j(\boldsymbol{r}\sigma). \tag{51}$$

Finally, for the RPA amplitudes given by non-hermitian matrix $\tilde{\rho}_{ij}$ one has

$$\psi_i(\boldsymbol{r}'\sigma') = \sum_j \tilde{\rho}_{ij}^*\phi_j^*(\boldsymbol{r}'\sigma'). \tag{52}$$

To derive the fields, first we recall that the variation of the non-local density, over the wave function reads

$$\begin{aligned} \delta\rho_{v\mu}(\boldsymbol{r}, \boldsymbol{r}') &= \sum_i \sum_{\sigma'} \frac{\partial\rho_{v\mu}(\boldsymbol{r}, \boldsymbol{r}')}{\partial\psi_i(\boldsymbol{r}'\sigma')}\delta\psi_i(\boldsymbol{r}'\sigma') \\ &= \sum_i \sum_{\sigma\sigma'} \langle\sigma'|\sigma_{v\mu}|\sigma\rangle\,\phi_i(\boldsymbol{r}\sigma)\delta\psi_i(\boldsymbol{r}'\sigma'). \end{aligned} \tag{53}$$

Therefore, variation of the primary density is given by

$$\delta\rho_{n'L'v'J'M'} = \sum_i \sum_{\sigma\sigma'} \left\{[K_{n'L'}\sigma_{v';\sigma'\sigma}]_{J'M'}\phi_i(\boldsymbol{r}\sigma)\delta\psi_i(\boldsymbol{r}'\sigma')\right\}_{\boldsymbol{r}'=\boldsymbol{r}}. \tag{54}$$

At this point, operators $K_{n'L'}$ mix derivatives acting on the variables $\boldsymbol{r}$ and $\boldsymbol{r}'$, cf. Eq. (18). By using the Wigner-Eckart theorem, we can express them as sums of products of derivatives $D_{mIM_I}$ and $D'_{m'I'M'_I}$, which act on $\boldsymbol{r}$ and $\boldsymbol{r}\,'$, respectively, that is,

$$K_{n'L'M'_L} \;=\; \sum_{mIm'I'} K^{n'L'}_{mIm'I'} \sum_{M_I M'_I} C^{L'M'_L}_{IM_I I'M'_I} D_{mIM_I} D'_{m'I'M'_I}, \tag{55}$$

where the order of derivative is conserved,

$$n' = m + m', \tag{56}$$

and the triangle rule of angular momentum coupling must be obeyed. Numerical coefficients $K^{n'L'}_{mIm'I'}$ can be calculated by using methods of symbolic programming. At N$^3$LO, only 91 coefficients $K^{n'L'}_{mIm'I'}$ are needed, so they can easily be precalculated and stored.

We now can insert expressions (54) and (55) into Eq. (45) and remove condition $\boldsymbol{r}' = \boldsymbol{r}$ by adding the integral over $\boldsymbol{r}\,'$ and the function $\delta(\boldsymbol{r} - \boldsymbol{r}\,')$, that is,

$$\begin{aligned} \delta\mathcal{E} \;&=\; \sum_{n'L'v'J'} \int \mathrm{d}^3\boldsymbol{r} \int \mathrm{d}^3\boldsymbol{r}\,'\delta(\boldsymbol{r} - \boldsymbol{r}\,') \\ &\times\; \sum_i \sum_\sigma [\tilde{U}_{n'L'v'J'}(\boldsymbol{r})[K_{n'L'}\sigma_{v';\sigma'\sigma}]_{J'}]_0 \phi_i(\boldsymbol{r}\sigma)\delta\psi_i(\boldsymbol{r}\,'\sigma'). \end{aligned} \tag{57}$$

This allows us to integrate by parts over $\boldsymbol{r}\,'$ and transfer the action of $D'_{m'I'M'_I}$ onto the delta function. With all the angular momenta couplings shown explicitly, this gives

$$\begin{aligned} \delta\mathcal{E} \;&=\; \sum_{n'L'v'J'} \int \mathrm{d}^3\boldsymbol{r} \int \mathrm{d}^3\boldsymbol{r}\,' \sum_{mIm'I'} \sum_{M_I M'_I} \left\{ D'^{T}_{m'I'M'_I}\delta(\boldsymbol{r} - \boldsymbol{r}\,') \right\} \\ &\times\; \sum_i \sum_{\sigma\sigma'} \sum_{M'_L\mu'} \sum_{M'} C^{J'M'}_{L'M'_L v'\mu'} \tfrac{(-1)^{J'-M'}}{\sqrt{2J'+1}} \tilde{U}_{n'L'v'J',-M'}(\boldsymbol{r}) \\ &\times\; K^{n'L'}_{mIm'I'} C^{L'M'_L}_{IM_I I'M'_I} D_{mIM_I} \sigma_{v'\mu';\sigma'\sigma} \phi_i(\boldsymbol{r}\sigma)\delta\psi_i(\boldsymbol{r}\,'\sigma'). \end{aligned} \tag{58}$$

The action of $D'^{T}_{m'I'M'_I}$ onto the delta function can be replaced by that of $(-1)^{m'} D^{T}_{m'I'M'_I}$, and then the integration by parts over $\boldsymbol{r}$ gives,

$$\delta\mathcal{E} \;=\; \sum_{n'L'v'J'} \int \mathrm{d}^3\boldsymbol{r} \int \mathrm{d}^3\boldsymbol{r}\,' \sum_{mIm'I'} \sum_{M_I M'_I} \delta(\boldsymbol{r} - \boldsymbol{r}\,')$$

$$\times \sum_i \sum_{\sigma\sigma'} \sum_{M_L'\mu'} \sum_{M'} C^{J'M'}_{L'M_L'v'\mu'} \frac{(-1)^{J'-M'}}{\sqrt{2J'+1}} (-1)^{m'} D_{m'I'M_I'} \tilde{U}_{n'L'v'J',-M'}(\boldsymbol{r})$$
$$\times K^{n'L'}_{mIm'I'} C^{L'M_L'}_{IM_I I'M_I'} D_{mIM_I} \sigma_{v'\mu';\sigma'\sigma} \phi_i(\boldsymbol{r}\sigma) \delta\psi_i(\boldsymbol{r}'\sigma'). \tag{59}$$

In the resulting local integral we request that the operator acting on $\phi_i(\boldsymbol{r}\sigma)$ is equal to the mean-field operator, which gives

$$h(\rho) = \sum_{n'L'v'J'} \sum_{mIm'I'} \sum_{M_I M_I'} \sum_{M_L'\mu'} \sum_{M'} C^{J'M'}_{L'M_L'v'\mu'} \frac{(-1)^{J'-M'}}{\sqrt{2J'+1}}$$
$$\times (-1)^{m'} K^{n'L'}_{mIm'I'} C^{L'M_L'}_{IM_I I'M_I'} D_{m'I'M_I'} \tilde{U}_{n'L'v'J',-M'}(\boldsymbol{r}) \sigma_{v'\mu'} D_{mIM_I}. \tag{60}$$

In this form, the mean-field operator is expressed through sums of derivative operators standing on both sides of the potentials $\tilde{U}_{n'L'v'J',-M'}$. This form can easily be used in the calculation of the matrix elements, because the left derivative operator can simply be applied onto the left wave-function through the integration by parts. Moreover, potentials $\tilde{U}_{n'L'v'J',-M'}$ are related to the secondary densities by very simple relations (46). However, it turns out that numerical calculations are much faster if the derivatives appear only on one side of the potentials, like it was postulated in Eq. (7), that is,

$$h(\rho) = \sum_{n'L'v'J'} [U_{n'L'v'J'}(\boldsymbol{r})[D_{n'L'}\sigma_{v'}]_{J'}]_0$$
$$= \sum_{n'L'v'J'} \sum_{M_L'\mu'} \frac{(-1)^{J'-M'}}{\sqrt{2J'+1}} C^{J'M'}_{L'M_L'v'\mu'} U_{n'L'v'J',-M'}(\boldsymbol{r}) \sigma_{v'\mu'} D_{n'L'M_L'}. \tag{61}$$

*3.4. Fields for terms containing additional density dependence*

In the Skyrme functional, a dependence on the isoscalar density is usually added to the four terms that are zero order in derivatives, $\rho_\tau^2$ and $\boldsymbol{s}_\tau^2$ ($\tau = 0, 1$), that is

$$T^{\alpha,0000}_{00,0000}(\tau) = C^{\alpha,0000}_{00,0000}(\tau)\, \rho_0^\alpha \rho_\tau^2, \tag{62}$$
$$T^{\alpha,0011}_{00,0011}(\tau) = C^{\alpha,0011}_{00,0011}(\tau)\, \rho_0^\alpha \boldsymbol{s}_\tau^2. \tag{63}$$

We use the extra index $\alpha$ on the coupling constant to distinguish the notation from the one used for the density-independent terms.

The contribution to the fields is obtained from the variation of the energy, which we show here only for the first of the above two terms, namely,

$$\begin{aligned}\frac{\delta\mathcal{E}^{\alpha,0000}_{00,0000}(\tau)}{\delta\phi^*_i(\tau')} &= \frac{\partial\left(C^{\alpha,0000}_{00,0000}(\tau)\,\rho_0^\alpha\right)}{\partial\phi^*_i(\tau')}\rho_\tau^2 + C^{\alpha,0000}_{00,0000}(\tau)\,\rho_0^\alpha\frac{\partial\rho_\tau^2}{\partial\phi^*_i(\tau')}\\ &= C^{\alpha,0000}_{00,0000}(\tau)\left[\alpha\rho_0^{\alpha-1}\rho_\tau^2\delta_{\tau',0} + 2\rho_0^\alpha\rho_\tau\delta_{\tau',\tau}\right]\phi_i(\tau')\end{aligned}$$

which gives the additional contributions

$$\sum_{\tau=0}^{1} C^{\alpha,0000}_{00,0000}(\tau)\left[\alpha\rho_0^{\alpha-1}\rho_\tau^2\delta_{\tau',0} + 2\rho_0^\alpha\rho_\tau\delta_{\tau',\tau}\right] \qquad (64)$$

to $U_{0000,0}(\tau')$.

*3.5. Rearrangement terms*

For density-dependent terms, the total energy obtained from the proton and neutron eigenvalues $\epsilon_i^{p,n}$ and kinetic energies $T^{p,n}$

$$E = \frac{1}{2}(T^p + T^n) + \frac{1}{2}\sum_i(\epsilon_i^p + \epsilon_i^n) + E_{RR} \qquad (65)$$

includes the additional rearrangement term

$$E_{RR} = \frac{1}{2}(T^p + T^n) - \frac{1}{2}\sum_i(\epsilon_i^p + \epsilon_i^n) + \int \mathrm{d}^3\boldsymbol{r}\,\mathcal{H}(\boldsymbol{r}). \qquad (66)$$

For spherical symmetry the terms with density-dependent coupling constants in the Skyrme functional only involve the $\rho_\tau$ densities and for these a straightforward derivation gives

$$E_{RR} = -\frac{1}{2}\int \mathrm{d}^3\boldsymbol{r}\,\left(\frac{\partial\mathcal{H}}{\partial\rho_0}\rho_0 + \frac{\partial\mathcal{H}}{\partial\rho_1}\rho_1 - 2\mathcal{H}\right). \qquad (67)$$

*3.6. Densities*

In order to calculate the potentials in Eq. (46) we have to determine all secondary densities:

$$\rho_{mI,nLvJ,J'M'}(\boldsymbol{r}) = [D_{mI}\rho_{nLvJ}(\boldsymbol{r})]_{J'M'} = [D_{mI}[K_{nL}\rho_v(\boldsymbol{r}_1,\boldsymbol{r}_2)]_J]_{J'M'}, \qquad (68)$$

where $D_{mI}$ is built by coupling gradients $\nabla_1+\nabla_2$ and $K_{nL}$ is built by coupling gradients $(\nabla_1-\nabla_2)/2i$. Here and below we understand that $\boldsymbol{r}=\boldsymbol{r}_1=\boldsymbol{r}_2$ is set after performing all the differentiations.

In analogy with Eq. (55), we can split these operators as

$$K_{nLM_L} = \sum_{rRr'R'} K^{nL}_{rRr'R'} \sum_{M_RM'_R} C^{LM_L}_{RM_RR'M'_R} D^{(1)}_{rRM_R} D^{(2)}_{r'R'M'_R}, \tag{69}$$

$$D_{mIM_I} = (2i)^m \sum_{pPp'P'} (-1)^{p'} K^{mI}_{pPp'P'} \sum_{M_PM'_P} C^{IM_I}_{PM_PP'M'_P} D^{(1)}_{pPM_P} D^{(2)}_{p'P'M'_P}, \tag{70}$$

where $n=r+r'$ and $m=p+p'$. Then the density (68) reads

$$\begin{aligned}\rho_{mI,nLvJ,J'M'}(\boldsymbol{r}) = &\sum_{M_IM_J} C^{J'M'}_{IM_IJM_J} \sum_{M_L\mu} C^{JM_J}_{LM_Lv\mu} \\ &\times (2i)^m \sum_{pPp'P'} (-1)^{p'} K^{mI}_{pPp'P'} \sum_{M_PM'_P} C^{IM_I}_{PM_PP'M'_P} D^{(1)}_{pPM_P} D^{(2)}_{p'P'M'_P} \\ &\times \sum_{rRr'R'} K^{nL}_{rRr'R'} \sum_{M_RM'_R} C^{LM_L}_{RM_RR'M'_R} D^{(1)}_{rRM_R} D^{(2)}_{r'R'M'_R} \rho_{v\mu}(\boldsymbol{r}_1,\boldsymbol{r}_2).\end{aligned} \tag{71}$$

We now introduce coefficients $D^{uU}_{rRpP}$, which allow for expressing products of derivatives as:

$$D^{(1)}_{rRM_R} D^{(1)}_{pPM_P} = \sum_{uUM_U} C^{UM_U}_{RM_RPM_P} D^{uU}_{rRpP} D^{(1)}_{uUM_U}, \tag{72}$$

$$D^{(2)}_{r'R'M'_R} D^{(2)}_{p'P'M'_P} = \sum_{u'U'M'_U} C^{U'M'_U}_{R'M'_RP'M'_P} D^{u'U'}_{r'R'p'P'} D^{(2)}_{u'U'M'_U}, \tag{73}$$

where $u=r+p$ and $u'=r'+p'$, that is, $u+u'=m+n$. These coefficients can be calculated by using methods outlined in A. At N$^3$LO, only 91 coefficients $D^{uU}_{rRpP}$ are needed, so they can easily be precalculated and stored. The sum of products of four Clebsh-Gordan coefficients can now be recoupled (see Eq. 8.7(20) in Ref. [6]) as:

$$\begin{aligned}&\sum_{M_PM'_P} \sum_{M_RM'_R} C^{IM_I}_{PM_PP'M'_P} C^{LM_L}_{RM_RR'M'_R} C^{UM_U}_{RM_RPM_P} C^{U'M'_U}_{R'M'_RP'M'_P} \\ &= \sqrt{(2I+1)(2L+1)(2U+1)(2U'+1)} \\ &\times \sum_{WM_W} \left\{ \begin{array}{ccc} P' & P & I \\ R' & R & L \\ U' & U & W \end{array} \right\} C^{WM_W}_{UM_UU'M'_U} C^{WM_W}_{LM_LIM_I}\end{aligned} \tag{74}$$

Subsequently, the sum of products of three Clebsh-Gordan coefficients can be recoupled (see Eq. 8.7(12) in Ref. [6]) as:

$$\sum_{M_J M_L M_I} C^{WM_W}_{LM_L IM_I} C^{J'M'}_{IM_I JM_J} C^{JM_J}_{LM_L v\mu} = (-1)^{W+v-J'} \sqrt{(2W+1)(2J+1)} \left\{ \begin{array}{ccc} L & I & W \\ J' & v & J \end{array} \right\} C^{J'M'}_{WM_W v\mu}. \quad (75)$$

The last two remaining Clebsh-Gordan coefficients can be absorbed in the following definition of the coupled derivative of the density:

$$\begin{aligned} \rho^{uUu'U'W}_{vJ'M'}(\boldsymbol{r}) &= [[D^{(1)}_{uU} D^{(2)}_{u'U'}]_W \rho_v(\boldsymbol{r}_1, \boldsymbol{r}_2)]_{J'M'} \quad (76) \\ &= \sum_{M_W\mu} C^{J'M'}_{WM_W v\mu} \sum_{M_U M'_U} C^{WM_W}_{UM_U U'M'_U} D^{(1)}_{uUM_U} D^{(2)}_{u'U'M'_U} \rho_{v\mu}(\boldsymbol{r}_1, \boldsymbol{r}_2), \end{aligned}$$

which finally gives

$$\rho_{mI,nLvJ,J'M'}(\boldsymbol{r}) = \sum_{uUu'U'W} A^{uUu'U',W}_{mI,nLvJ,J'} \rho^{uUu'U'W}_{vJ'M'}(\boldsymbol{r}), \quad (77)$$

where coefficients $A^{uUu'U',W}_{mI,nLvJ,J'}$ result from summing up all intrinsic indices:

$$\begin{aligned} A^{uUu'U',W}_{mI,nLvJ,J'} &= (2i)^m \sum_{pPp'P'} \sum_{rRr'R'} (-1)^{p'} K^{mI}_{pPp'P'} K^{nL}_{rRr'R'} D^{uU}_{rRpP} D^{u'U'}_{r'R'p'P'} \\ &\times \sqrt{(2I+1)(2L+1)(2U+1)(2U'+1)} \left\{ \begin{array}{ccc} P' & P & I \\ R' & R & L \\ U' & U & W \end{array} \right\} \\ &\times (-1)^{W+v-J'} \sqrt{(2W+1)(2J+1)} \left\{ \begin{array}{ccc} L & I & W \\ J' & v & J \end{array} \right\}. \quad (78) \end{aligned}$$

At $N^3LO$, only 3138 coefficients $A^{uUu'U',W}_{mI,nLvJ,J'}$ are needed, so they can easily be precalculated and stored.

## 4. The $N^3LO$ potentials, fields, and densities in the spherical harmonic oscillator basis

### *4.1. Spherical HO basis*

The standard spherical HO wave functions, are given by

$$\phi_{Nljm}(r,\theta,\phi,\sigma) = b^{3/2} F_{Nl}(br) e^{-\frac{1}{2}(br)^2} \sum_{m_l m_s} C^{jm}_{lm_l \frac{1}{2} m_s} Y_{lm_l}(\theta,\phi) \chi_{\frac{1}{2} m_s}(\sigma), (79)$$

where $b$ is the oscillator constant,

$$b = \sqrt{\frac{m\omega}{\hbar}}, \tag{80}$$

and $F_{Nl}(br) =$ are proportional to the standard Laguerre polynomials [7].

To calculate the secondary densities (76), one has to act on the space part of the wave function (79) with the derivative operators $D_{nLM_L}$. This leads to defining the polynomials $F^{km_k}_{nLM_LNlm_l}(br)$ such that

$$\begin{aligned} & D_{nLM_L} b^{3/2} F_{Nl}(br) e^{-\frac{1}{2}(br)^2} Y_{lm_l}(\theta,\phi) \\ & = b^{n+3/2} \sum_{km_k} F^{km_k}_{nLM_LNlm_l}(br) e^{-\frac{1}{2}(br)^2} Y_{km_k}(\theta,\phi), \end{aligned} \tag{81}$$

where $m_k = M_L + m_l$. From the Wigner-Eckart theorem, these polynomials must have the form:

$$F^{km_k}_{nLM_LNlm_l}(br) \;\; = \;\; \tfrac{1}{\sqrt{2k+1}} C^{km_k}_{lm_lLM_L} F^k_{nLNl}(br), \tag{82}$$

and our goal is to determine the set of reduced polynomials $F^k_{nLNl}(br)$, in terms of which the derivatives of the spherical HO wave functions read

$$\begin{aligned} D_{nLM_L}\phi_{Nljm}(r,\theta,\phi,\sigma) \;\; &= \;\; b^{n+3/2} \sum_{km_k} \sum_{m_lm_s} \tfrac{1}{\sqrt{2k+1}} C^{km_k}_{lm_lLM_L} C^{jm}_{lm_l\frac{1}{2}m_s} \\ & \times \;\; F^k_{nLNl}(br) e^{-\frac{1}{2}(br)^2} Y_{km_k}(\theta,\phi) \chi_{\frac{1}{2}m_s}(\sigma), \\ & = \;\; -b^{n+3/2}\sqrt{2j+1} \sum_{km_k} \sum_{m_s} \sum_{im_i} C^{im_i}_{LM_Ljm} C^{im_i}_{\frac{1}{2}m_skm_k} \\ & \times \;\; \left\{ \begin{array}{ccc} l & L & k \\ i & \frac{1}{2} & j \end{array} \right\} F^k_{nLNl}(br) e^{-\frac{1}{2}(br)^2} Y_{km_k}(\theta,\phi) \chi_{\frac{1}{2}m_s}(\sigma). \end{aligned} \tag{83}$$

Explicit form of $F^k_{nLNl}(br)$ can be calculated by using the Wigner-Eckart theorem again, namely, Eq. (81) must have the form

$$D_{nLM_L}\phi_{Nlm_l} \;\; = \;\; \sum_{N'km_k} \tfrac{1}{\sqrt{2k+1}} C^{km_k}_{lm_lLM_L} \langle \phi_{N'k} || D_{nL} || \phi_{Nl} \rangle \phi_{N'km_k}, \tag{84}$$

where $\phi_{Nlm_l}$ is the space part of the wave function (79). Then we have polynomials $F^k_{nLNl}(br)$ expressed through the Laguerre polynomials as:

$$b^n F^k_{nLNl}(br) = \sum_{N'} \langle \phi_{N'k} || D_{nL} || \phi_{Nl} \rangle F_{N'k}(br), \tag{85}$$

where the reduced matrix element can be calculated by considering only one matrix element, namely,

$$\langle \phi_{N'k,m_k=0} | D_{nL,M_L=0} | \phi_{Nl,m_l=0} \rangle = \tfrac{1}{\sqrt{2k+1}} C^{k0}_{l0L0} \langle \phi_{N'k} || D_{nL} || \phi_{Nl} \rangle. \tag{86}$$

Note that the Clebsh-Gordan coefficient $C^{k0}_{l0L0}$ is not zero for angular momenta restricted by the parity conservation, $(-1)^{L+l-k} = 1$.

The parity conservation induces specific conditions on the polynomials $F^k_{nLNl}(br)$. Indeed, by comparing parities of both sides of Eq. (81) we see that

$$F^k_{nLNl}(br) = 0, \quad \text{for} \quad (-1)^{n+l+k} = -1. \tag{87}$$

Equivalently, since for all derivative operators we have $(-1)^{n+L} = 1$, we see that

$$F^k_{nLNl}(br) = 0, \quad \text{for} \quad (-1)^{L+l+k} = -1. \tag{88}$$

Since polynomials $F_{Nl}(br)$ are real, phases of polynomials $F^k_{nLNl}(br)$ are fixed by those of the derivative operators (31) and spherical harmonics [6],

$$Y^*_{JM}(\theta, \phi) = (-1)^{-M} Y_{J,-M}(\theta, \phi), \tag{89}$$

that is,

$$F^{k*}_{nLNl}(br) = (-1)^{l-k} F^k_{nLNl}(br) = (-1)^n F^k_{nLNl}(br) = (-1)^L F^k_{nLNl}(br), \tag{90}$$

where the last two equivalent forms result from Eqs (87) and (88).

The $F^k_{nLNl}$ polynomials are calculated using formulas for spherical derivatives (see [6]) combined with recursion relations for derivatives of Laguerre polynomials. In this way spherical derivatives of one of the basis functions $\phi_{Nljm}(\boldsymbol{r}, \sigma) = g_{nl}(r) C^{jm}_{lm_l, \frac{1}{2}\sigma} Y_{lm}(\theta, \phi)$ can be expressed as a sum of functions

$$\begin{aligned}\nabla_{1\mu_1} .. \nabla_{1\mu_N} g_{nl}(r) Y_{lm} &= \sum_{i_1,..,i_N=-1}^{1} [a(l, n, i_1, .., i_N, r) g_{n-1,l}(r) \\ &+ b(l, n, i_1, .., i_N, r) g_{nl}(r)] Y_{l+i_1+..+i_n, m+\mu_1+..+\mu_N},\end{aligned}$$

where the $a$ and $b$ coefficients needed for the different orders were derived using symbolic programming. In this way we obtain analytical expressions for all derivatives which can then be calculated with good accuracy.

*4.2. Densities in the spherical HO basis*

For any density matrix one can always perform a multipole expansion. This strategy fits very well our applications to the spherical HF solutions, where only the monopole component of the density matrix is nonzero, and to the RPA applications in spherical nuclei, where all multipole excitations separate from one another. Therefore, in what follows we consider the density matrix of multipolarity $J$ in the form given by the Wigner-Eckart theorem:

$$\tilde{\rho}^{JM}_{Nljm,N'l'j'm'} = \frac{1}{\sqrt{2j+1}} C^{jm}_{j'm'JM} \langle \phi_{Nlj} || \tilde{\rho}^{J} || \phi_{N'l'j'} \rangle, \tag{91}$$

and depending on its reduced matrix elements $\langle \phi_{Nlj} || \tilde{\rho}^{J} || \phi_{N'l'j'} \rangle$ Then, the non-local densities can be expressed in terms of the spherical HO wave functions (79) as

$$\tilde{\rho}^{J''M''}_{v\mu}(\boldsymbol{r}_1, \boldsymbol{r}_2) = \sum_{\substack{Nljm\sigma \\ N'l'j'm'\sigma'}} \phi_{Nljm}(\boldsymbol{r}_1, \sigma) \tilde{\rho}^{J''M''}_{Nljm,N'l'j'm'} \phi^*_{N'l'j'm'}(\boldsymbol{r}_2, \sigma') \langle \sigma' | \sigma_{v\mu} | \sigma \rangle, \tag{92}$$

that is,

$$\begin{aligned}
\tilde{\rho}^{J''M''}_{v\mu}(\boldsymbol{r}_1, \boldsymbol{r}_2) = & \sum_{\substack{Nljm\sigma \\ N'l'j'm'\sigma'}} b^3 e^{-\frac{1}{2}(br_1)^2 - \frac{1}{2}(br_2)^2} \langle \sigma' | \sigma_{v\mu} | \sigma \rangle \\
\times & \; F_{Nl}(br_1) \sum_{m_l m_s} C^{jm}_{lm_l \frac{1}{2} m_s} Y_{lm_l}(\theta_1, \phi_1) \chi_{\frac{1}{2} m_s}(\sigma) \\
\times & \; \tfrac{1}{\sqrt{2j+1}} C^{jm}_{j'm'J''M''} \langle \phi_{Nlj} || \tilde{\rho}^{J''} || \phi_{N'l'j'} \rangle \\
\times & \; F_{N'l'}(br_2) \sum_{m'_l m'_s} C^{j'm'}_{l'm'_l \frac{1}{2} m'_s} Y^*_{l'm'_l}(\theta_2, \phi_2) \chi_{\frac{1}{2} m'_s}(\sigma').
\end{aligned} \tag{93}$$

We can now replace the spin coordinates by the spin projections,

$$\chi_{\frac{1}{2} m_s}(\sigma) = \delta_{\frac{1}{2} m_s, \sigma} \quad , \quad \chi_{\frac{1}{2} m'_s}(\sigma') = \delta_{\frac{1}{2} m'_s, \sigma'}, \tag{94}$$

and we can use the Wigner-Eckart theorem for the Pauli matrices,

$$\langle\tfrac{1}{2}m_s'|\sigma_{v\mu}|\tfrac{1}{2}m_s\rangle = \tfrac{1}{\sqrt{2}}C^{\frac{1}{2}m_s'}_{\frac{1}{2}m_s v\mu}\langle\tfrac{1}{2}||\sigma_v||\tfrac{1}{2}\rangle, \tag{95}$$

with

$$\langle\tfrac{1}{2}||\sigma_0||\tfrac{1}{2}\rangle = \sqrt{2} \quad , \quad \langle\tfrac{1}{2}||\sigma_1||\tfrac{1}{2}\rangle = -i\sqrt{6}. \tag{96}$$

After inserting the nonlocal density (93) into the expression for local densities (76), and after acting with derivatives on spherical wave functions, as in Eqs. (81) and (82), we obtain

$$\begin{aligned}\tilde{\rho}^{uUu'U'W,J''M''}_{vJ'M'}(\boldsymbol{r}) = &\sum_{M_W\mu} C^{J'M'}_{WM_Wv\mu}\sum_{M_UM_U'} C^{WM_W}_{UM_UU'M_U'}(-1)^{U'-M_U'}\\ &\times \sum_{\substack{Nljmm_s\\ N'l'j'm'm_s'}} b^{3+u+u'}e^{-(br)^2}\tfrac{1}{\sqrt{2}}C^{\frac{1}{2}m_s'}_{\frac{1}{2}m_sv\mu}\langle\tfrac{1}{2}||\sigma_v||\tfrac{1}{2}\rangle\\ &\times \sum_{km_km_l}\tfrac{1}{\sqrt{2k+1}}C^{km_k}_{lm_lUM_U}F^k_{uUNl}(br)C^{jm}_{lm_l\frac{1}{2}m_s}Y_{km_k}(\theta,\phi)\\ &\times \tfrac{1}{\sqrt{2j+1}}C^{jm}_{j'm'J''M''}\langle\phi_{Nlj}||\tilde{\rho}^{J''}||\phi_{N'l'j'}\rangle\\ &\times \sum_{k'm_k'm_l'}\tfrac{1}{\sqrt{2k'+1}}C^{k'm_k'}_{l'm_l'U',-M_U'}F^{k'*}_{u'U'N'l'}(br)C^{j'm'}_{l'm_l'\frac{1}{2}m_s'}Y^*_{k'm_k'}(\theta,\phi),\end{aligned} \tag{97}$$

where, by using the phase convention $D^{(2)}_{u'U'M_U'} = (-1)^{U'-M_U'}D^{(2)*}_{u'U',-M_U'}$, we have introduced the complex conjugation into the derivative operator (31).

After a lengthy but straightforward derivation presented in B, we obtain the following result:

$$\tilde{\rho}^{uUu'U'W,J''M''}_{vJ'M'}(\boldsymbol{r}) = \sum_{TM_T}\tilde{\rho}^{uUu'U'W,J''}_{vTJ'}(br)e^{-(br)^2}C^{TM_T}_{J'M'J''M''}Y_{TM_T}(\theta,\phi), \tag{98}$$

for the radial form factors $\tilde{\rho}^{uUu'U'W,J''}_{vTJ'}(br)$ given by:

$$\begin{aligned}\tilde{\rho}^{uUu'U'W,J''}_{vTJ'}(br) = &(-1)^{1+v}\langle\tfrac{1}{2}||\sigma_v||\tfrac{1}{2}\rangle\sqrt{\tfrac{(2J'+1)}{4\pi}}\sum_{\substack{Nljk\\ N'l'j'k'}}(-1)^{j'+j}B^{ljkU,WJ''}_{l'j'k'U',vTJ'}\\ &\times b^{3+u+u'}F^k_{uUNl}(br)\langle\phi_{Nlj}||\tilde{\rho}^{J''}||\phi_{N'l'j'}\rangle F^{k'}_{u'U'N'l'}(br),\end{aligned} \tag{99}$$

where

$$B^{ljkU,WJ''}_{l'j'k'U',vTJ'} = (-1)^k(-1)^T(-1)^{U-W}C^{T0}_{k0k'0}\sqrt{\tfrac{(2k+1)(2k'+1)(2W+1)(2j+1)(2j'+1)}{(2T+1)}}$$
$$\times \sum_{T'}(-1)^{T'}(2T'+1)$$
$$\times \left\{\begin{array}{ccc} j & j' & J'' \\ \frac{1}{2} & \frac{1}{2} & v \\ l & l' & T' \end{array}\right\}\left\{\begin{array}{ccc} l & k & U \\ l' & k' & U' \\ T' & T & W \end{array}\right\}\left\{\begin{array}{ccc} v & T' & J'' \\ T & J' & W \end{array}\right\}. \quad (100)$$

In view of the fact that only the coefficients $B^{ljkU,WJ''}_{l'j'k'U',vTJ'}$ for $(-1)^{l+k+U} = 1$ and $(-1)^{l'+k'+U'} = 1$ are required in Eq. (99), we may replace them by coefficients $\tilde{B}^{ljkU,WJ''}_{l'j'k'U',vTJ'}$:

$$\tilde{B}^{ljkU,WJ''}_{l'j'k'U',vTJ'} = (-1)^l C^{T0}_{k0k'0}\sqrt{\tfrac{(2k+1)(2k'+1)(2W+1)(2j+1)(2j'+1)}{(2T+1)}}$$
$$\times \sum_{T'}(2T'+1)$$
$$\times \left\{\begin{array}{ccc} \frac{1}{2} & \frac{1}{2} & v \\ j' & j & J'' \\ l' & l & T' \end{array}\right\}\left\{\begin{array}{ccc} l' & l & T' \\ k' & k & T \\ U' & U & W \end{array}\right\}\left\{\begin{array}{ccc} v & T' & J'' \\ T & J' & W \end{array}\right\}, \quad (101)$$

where we have used symmetry properties of 9j symbols under the transposition of rows and columns and transposition with respect to the main diagonal.

Finally, all secondary densities of Eq. (77) can now be calculated in terms of one compact expression:

$$\tilde{\rho}^{J''M''}_{mI,nLvJ,J'M'}(\boldsymbol{r}) = \sum_{TM_T}\tilde{\rho}^{TJ''}_{mI,nLvJ,J'}(br)e^{-(br)^2}C^{TM_T}_{J'M'J''M''}Y_{TM_T}(\theta,\phi), \quad (102)$$

where the radial form factors read

$$\tilde{\rho}^{TJ''}_{mI,nLvJ,J'}(br) = \sum_{uUu'U'W} A^{uUu'U'W}_{mI,nLvJ,J'}\,\tilde{\rho}^{uUu'U'W,J''}_{vTJ'}(br). \quad (103)$$

*4.3. Potentials in the spherical HO basis*

In the spherical basis, the secondary densities to be used in Eq. (46) have the form given in Eq. (102). Therefore, the potentials in Eq. (60) acquire the

form

$$\tilde{U}^{J''M''}_{n'L'v'J'M'}(\boldsymbol{r}) = \sum_{TM_T} \tilde{U}^{TJ''}_{n'L'v'J'}(br) e^{-(br)^2} C^{TM_T}_{J'M'J''M''} Y_{TM_T}(\theta,\phi), \tag{104}$$

where the radial form factors read:

$$\tilde{U}^{TJ''}_{n'L'v'J'}(br) = \sum_{mI,nLvJ} \left\{ C^{n'L'v'J'}_{mI,nLvJ} + (-1)^{J-J'} C^{nLvJ}_{mI,n'L'v'J'} \right\} \tilde{\rho}^{TJ''}_{mI,nLvJ,J'}(br) \tag{105}$$

Similarly, potentials $U^{J''M''}_{n'L'v'J'M'}(\boldsymbol{r})$ in Eq. (61) read

$$U^{J''M''}_{n'L'v'J'M'}(\boldsymbol{r}) = \sum_{TM_T} U^{TJ''}_{n'L'v'J'}(br) e^{-(br)^2} C^{TM_T}_{J'M'J''M''} Y_{TM_T}(\theta,\phi), \tag{106}$$

and the radial from factors $U^{TJ''}_{n'L'v'J'}(br)$ can be calculated in the complete analogy to the results outlined in Section 2, see Eq. (8), namely,

$$U^{TJ''}_{n'L'v'J'}(br) = \sum_{a\alpha\beta;mI,nLvJ} C^{\beta}_{a,\alpha} \chi^{\beta;mI,nLvJ}_{a,\alpha;n'L'v'J'} \tilde{\rho}^{TJ''}_{mI,nLvJ,J'}(br). \tag{107}$$

*4.4. Matrix elements of the Hamiltonian (60) in the spherical HO basis*

In the spherical HO basis of Eq. (79), we can calculate the matrix elements of the single-particle Hamiltonian (60) in the following way:

$$\langle N'l'j'm'_j|\tilde{h}(\tilde{\rho}^{J''M''})|Nljm_j\rangle = \sum_{n'L'v'J'} \sum_{mIm'I'} \sum_{M_IM'_I} \sum_{M'_L\mu'} \int r^2 \mathrm{d}\,r \int \mathrm{d}\Omega \sum_{\sigma'\sigma}$$
$$\sum_{M'} C^{J'M'}_{L'M'_Lv'\mu'} \tfrac{(-1)^{J'-M'}}{\sqrt{2J'+1}} (-1)^{m'} K^{n'L'}_{mIm'I'} C^{L'M'_L}_{IM_II'M'_I} \langle\sigma'|\sigma_{v'\mu'}|\sigma\rangle$$
$$\phi^*_{N'l'j'm'_j}(r,\theta,\phi,\sigma')\, D_{m'I'M'_I} \tilde{U}^{J''M''}_{n'L'v'J',-M'}(\boldsymbol{r}) D_{mIM_I}\, \phi_{Nljm_j}(r,\theta,\phi,\sigma), \tag{108}$$

where $\tilde{h}(\tilde{\rho}^{J''M''})$ denotes the field calculated for the one-multipole density matrix $\tilde{\rho}^{J''M''}$ (91). The integration by parts now gives

$$\langle N'l'j'm'_j|\tilde{h}(\tilde{\rho}^{J''M''})|Nljm_j\rangle = \sum_{n'L'v'J'} \sum_{mIm'I'} \sum_{M_IM'_I} \sum_{M'_L\mu'} \int r^2 \mathrm{d}\,r \int \mathrm{d}\Omega \sum_{\sigma'\sigma}$$
$$\sum_{M'} C^{J'M'}_{L'M'_Lv'\mu'} \tfrac{(-1)^{J'-M'}}{\sqrt{2J'+1}} (-1)^{m'} K^{n'L'}_{mIm'I'} C^{L'M'_L}_{IM_II'M'_I} \langle\sigma'|\sigma_{v'\mu'}|\sigma\rangle (-1)^{m'+I'-M'_I}$$
$$\tilde{U}^{J''M''}_{n'L'v'J',-M'}(\boldsymbol{r}) \left(D_{m'I',-M'_I}\phi_{N'l'j'm'_j}(r,\theta,\phi,\sigma')\right)^* \left(D_{mIM_I}\phi_{Nljm_j}(r,\theta,\phi,\sigma)\right), \tag{109}$$

where we have used the hermitian-conjugation property (32) of the differential operators.

By inserting potentials (104) and derivatives of spherical wavefunctions (83) we have:

$$
\begin{aligned}
\langle N'l'j'm_j'|\tilde{h}(\tilde{\rho}^{J''M''})|Nljm_j\rangle &= \sum_{n'L'v'J'}\sum_{mIm'I'}\sum_{M_IM_I'}\sum_{M_L'\mu'}\int r^2 \mathrm{d}\, r\int \mathrm{d}\Omega\sum_{\sigma'\sigma} \\
&\sum_{M'} C^{J'M'}_{L'M_L'v'\mu'}\tfrac{(-1)^{J'-M'}}{\sqrt{2J'+1}}(-1)^{m'}K^{n'L'}_{mIm'I'}C^{L'M_L'}_{IM_II'M_I'}\langle\sigma'|\sigma_{v'\mu'}|\sigma\rangle(-1)^{m'+I'-M_I'} \\
&\times\sum_{TM_T}\tilde{U}^{TJ''}_{n'L'v'J'}(br)e^{-(br)^2}C^{TM_T}_{J',-M'J''M''}Y_{TM_T}(\theta,\phi) \\
&\times b^{m'+3/2}\sum_{k'm_k'}\sum_{m_l'm_s'}\tfrac{1}{\sqrt{2k'+1}}C^{k'm_k'}_{l'm_l'I',-M_I'}C^{j'm_j'}_{l'm_l'\frac{1}{2}m_s'} \\
&\times F^{k'*}_{m'I'N'l'}(br)e^{-\frac{1}{2}(br)^2}Y^*_{k'm_k'}(\theta,\phi)\chi_{\frac{1}{2}m_s'}(\sigma') \\
&\times b^{m+3/2}\sum_{km_k}\sum_{m_lm_s}\tfrac{1}{\sqrt{2k+1}}C^{km_k}_{lm_lIM_I}C^{jm_j}_{lm_l\frac{1}{2}m_s} \\
&\times F^{k}_{mINl}(br)e^{-\frac{1}{2}(br)^2}Y_{km_k}(\theta,\phi)\chi_{\frac{1}{2}m_s}(\sigma),
\end{aligned}
\tag{110}
$$

The angular part can be integrated explicitly, by using the multiplication law of spherical harmonics (Eq. 5.6(9) in Ref. [6]), and summations over $\sigma$ and $\sigma'$ can be performed as in Eqs. (94)–(96). This gives

$$
\begin{aligned}
\langle N'l'j'm_j'|\tilde{h}(\tilde{\rho}^{J''M''})|Nljm_j\rangle &= \sum_{n'L'v'J'}\sum_{mIm'I'}\sum_{M_IM_I'}\sum_{M_L'\mu'}\int r^2 \mathrm{d}\, r \\
&\sum_{M'} C^{J'M'}_{L'M_L'v'\mu'}\tfrac{(-1)^{J'-M'}}{\sqrt{2J'+1}}(-1)^{m'}K^{n'L'}_{mIm'I'}C^{L'M_L'}_{IM_II'M_I'}(-1)^{m'+I'-M_I'} \\
\times\; &\sum_{TM_T}\sum_{k'm_k'}\sum_{m_l'm_s'}\sum_{km_k}\sum_{m_lm_s}\tfrac{1}{\sqrt{2}}C^{\frac{1}{2}m_s'}_{\frac{1}{2}m_sv'\mu'}\langle\tfrac{1}{2}||\sigma_{v'}||\tfrac{1}{2}\rangle\sqrt{\tfrac{(2T+1)(2k+1)}{4\pi(2k'+1)}}C^{k'0}_{T0k0}C^{k'm_k'}_{TM_Tkm_k} \\
\times\; & b^{3+n'}F^{k'*}_{m'I'N'l'}(br)\tilde{U}^{TJ''}_{n'L'v'J'}(br)F^{k}_{mINl}(br)e^{-2(br)^2} \\
\times\; & C^{TM_T}_{J',-M'J''M''}\tfrac{1}{\sqrt{2k'+1}}C^{k'm_k'}_{l'm_l'I',-M_I'}C^{j'm_j'}_{l'm_l'\frac{1}{2}m_s'}\tfrac{1}{\sqrt{2k+1}}C^{km_k}_{lm_lIM_I}C^{jm_j}_{lm_l\frac{1}{2}m_s},
\end{aligned}
\tag{111}
$$

where we have used condition (56).

We may now proceed by calculating the reduced matrix element of the field:

$$\langle \phi_{N'l'j'} || \tilde{h}^{I''}(\tilde{\rho}^{J''M''}) || \phi_{Nlj} \rangle = \sum_{m_j m'_j M''_I} \frac{1}{\sqrt{2j'+1}} C^{j'm'_j}_{jm_j I''M''_I} \langle N'l'j'm'_j | \tilde{h}(\tilde{\rho}^{J''M''}) | Nljm_j \rangle. \tag{112}$$

Again, after a lengthy but straightforward derivation presented in C, we obtain the following result:

$$\begin{aligned} \langle \phi_{N'l'j'} || \tilde{h}^{I''}(\tilde{\rho}^{J''M''}) || \phi_{Nlj} \rangle &= \delta_{I''J''}(-1)^{J''} \\ \times & \sum_{n'L'v'J'} \sum_{mIm'I'} K^{n'L'}_{mIm'I'} (-1)^{v'} \langle \tfrac{1}{2} || \sigma_{v'} || \tfrac{1}{2} \rangle \sqrt{\tfrac{1}{4\pi}} \sum_{Tkk'} (-1)^T (2T+1) \\ \times & \int r^2 \mathrm{d}\, r e^{-2(br)^2} b^{3+n'} F^{k'}_{m'I'N'l'}(br) \tilde{U}^{TJ''}_{n'L'v'J'}(br) F^{k}_{mINl}(br) \\ \times & \; B^{ljkI,L'J''}_{l'j'k'I',v'TJ'}, \end{aligned} \tag{113}$$

where we see the same numerical coefficients (100) that already appeared in Eq. (99).

*4.5. Matrix elements of the Hamiltonian (7) in the spherical HO basis*

For the Hamiltonian in the form given by Eq. (7), a similar derivation gives the matrix elements in the spherical HO basis that can be written as:

$$\begin{aligned} \langle \phi_{N'l'j'} || \tilde{h}^{I''}(\tilde{\rho}^{J''M''}) || \phi_{Nlj} \rangle &= \delta_{I''J''} \\ \times & \int r^2 \mathrm{d}\, r b^{3+n'} e^{-2(br)^2} F^{l'}_{00N'l'}(br) \sum_{n'L'k} F^{k}_{n'L'Nl}(br) \\ \times & \sum_{v'J'T} U^{TJ''}_{n'L'v'J'}(br) \langle \tfrac{1}{2} || \sigma_{v'} || \tfrac{1}{2} \rangle h^{ljkL',J'J''}_{l'j',v'T} \end{aligned} \tag{114}$$

with the reduced form factors $U^{TJ''}_{n'L'v'J'}(br)$ given in Eq. (107), and

$$\begin{aligned} h^{ljkL',J'J''}_{l'j',v'T} &= \frac{1}{2\sqrt{\pi}} (-1)^{J''+l+v'+l'+L'} \frac{\sqrt{(2k+1)(2j'+1)(2j+1)}}{\sqrt{(2l'+1)}} \\ &\times (-1)^T (2T+1) C^{l'0}_{k0T0} \\ &\times \sum_{T'} (-1)^{T'} (2T'+1) \begin{Bmatrix} \frac{1}{2} & \frac{1}{2} & v' \\ j' & j & J'' \\ l' & l & T' \end{Bmatrix} \end{aligned}$$

$$\times \left\{ \begin{array}{ccc} T' & l & l' \\ k & T & L' \end{array} \right\} \left\{ \begin{array}{ccc} J'' & T & J' \\ L' & v' & T' \end{array} \right\} . \tag{115}$$

*4.6. The total potential energy in the spherical symmetry*

The energy related to one term in the energy density (38) reads

$$\mathcal{E}^{n'L'v'J'}_{mI,nLvJ} = \int \mathrm{d}^3\boldsymbol{r}\, C^{n'L'v'J'}_{mI,nLvJ} \sum_{M'} \tfrac{(-1)^{J'-M'}}{\sqrt{2J'+1}} \left(\rho^{J_2''M_2''}_{00,n'L'v'J',J'M'}\right) \left(\rho^{J_1''M_1''}_{mI,nLvJ,J',-M'}\right) . \tag{116}$$

We carry out the derivation for arbitrary values of multipole components, that is for $J_2''M_2'' \neq J_1''M_1'' \neq 00$, while the standard HF total energy in spherical symmetry corresponds to the particular case of $J_2''M_2'' = J_1''M_1'' = 00$. In the general case, the result corresponds to the energy matrix element between two multipole components, and allows one to calculated the total energy for the density matrix being a linear combination of multipole components, that is, for a deformed state.

After using the definition of reduced density in Eq. (102), the energy (116) becomes equal to:

$$\begin{aligned} \mathcal{E}^{n'L'v'J'}_{mI,nLvJ} = {} & \int \mathrm{d}^3\boldsymbol{r}\, C^{n'L'v'J'}_{mI,nLvJ} e^{-2(br)^2} \sum_{M'} \tfrac{(-1)^{J'-M'}}{\sqrt{2J'+1}} \\ & \times \sum_{T'M_T'} \tilde{\rho}^{T'J_2''}_{00,n'L'v'J',J'}(br) C^{T'M_T'}_{J'M',J_2''M_2''} Y_{T'M_T'}(\theta,\phi) \\ & \times \sum_{TM_T} \tilde{\rho}^{TJ_1''}_{mI,nLvJ,J'}(br) C^{TM_T}_{J',-M',J_1''M_1''} Y_{TM_T}(\theta,\phi) . \end{aligned}$$

By using the multiplication theorem for the spherical harmonics, one obtains the expression

$$\begin{aligned} \mathcal{E}^{n'L'v'J'}_{mI,nLvJ} &= \int \mathrm{d}^3\boldsymbol{r}\, C^{n'L'v'J'}_{mI,nLvJ} e^{-2(br)^2} \sum_{TT'Q} \tilde{\rho}^{T'J_2''}_{00,n'L'v'J',J'}(br)\, \tilde{\rho}^{TJ_1''}_{mI,nLvJ,J'}(br) \\ &\times \; C^{Q0}_{T0,T'0} \sqrt{\frac{(2T+1)(2T'+1)}{4\pi(2Q+1)}} \sum_{M'M_T'M_TM_Q} \tfrac{(-1)^{J'-M'}}{\sqrt{2J'+1}} C^{T'M_T'}_{J'M',J_2''M_2''} \\ &\times \; C^{TM_T}_{J',-M',J_1''M_1''} C^{QM_Q}_{TM_T,T'M_T'} Y_{QM_Q}(\theta,\phi) , \end{aligned} \tag{117}$$

which, for density-independent coupling constants, can be integrated over $\theta$ and $\phi$. After summing up the Clebsh-Gordan coefficients, one obtains:

$$\begin{aligned}\mathcal{E}_{mI,nLvJ}^{n'L'v'J'} &= \frac{(-1)^{J_1''-M_1''}}{\sqrt{2J'+1}(2J_1''+1)}\delta_{J_2''J_1''}\delta_{M_2'',-M_1''}\int r^2 \mathrm{d}\, r\, C_{mI,nLvJ}^{n'L'v'J'} e^{-2(br)^2} \\ &\times \sum_T (-1)^T (2T+1) \tilde{\rho}_{00,n'L'v'J',J'}^{TJ_2''}(br)\, \tilde{\rho}_{mI,nLvJ,J'}^{TJ_1''}(br)\,. \quad (118)\end{aligned}$$

One thus obtains the correct result that *for density-independent coupling constants*, the potential energy is diagonal in different multipoles.

When the coupling constants do depend on density, the angular integral cannot be performed, and a similar derivation gives the general result:

$$\begin{aligned}\mathcal{E}_{mI,nLvJ}^{n'L'v'J'} &= \int \mathrm{d}^3\boldsymbol{r}\, C_{mI,nLvJ}^{n'L'v'J'} e^{-2(br)^2} \sum_{TT'Q} \tilde{\rho}_{00,n'L'v'J',J'}^{T'J_2''}(br)\, \tilde{\rho}_{mI,nLvJ,J'}^{TJ_1''}(br) \\ &\times C_{T0,T'0}^{Q0} \frac{(2T+1)(2T'+1)}{\sqrt{4\pi}} \\ &\times \sum_{M_Q} C_{J_1''M_1'',J_2''M_2''}^{QM_Q} \left\{ \begin{array}{ccc} J' & J' & 0 \\ T' & T & Q \\ J_2'' & J_1'' & Q \end{array} \right\} Y_{QM_Q}(\theta,\phi)\,. \quad (119)\end{aligned}$$

For the spherical case, one has $J_1'' = J_2'' = 0$, which implies $T = J'$, and both expressions (118) and (119) reduce to:

$$\begin{aligned}\mathcal{E}_{mI,nLvJ}^{n'L'v'J'} &= (-1)^{J'}\sqrt{2J'+1}\int r^2 \mathrm{d}\, r\, C_{mI,nLvJ}^{n'L'v'J'} e^{-2(br)^2} \\ &\times \tilde{\rho}_{00,n'L'v'J',J'}^{J'0}(br)\, \tilde{\rho}_{mI,nLvJ,J'}^{J'0}(br)\,. \quad (120)\end{aligned}$$

For density-dependent terms, such as those in Eqs. (62) and (63), the total energy reads

$$\begin{aligned}\mathcal{E}_{mI,nLvJ}^{\alpha,n'L'v'J'} &= (-1)^{J'}\sqrt{2J'+1}C_{mI,nLvJ}^{\alpha,n'L'v'J'}\int r^2 \mathrm{d}\, r\, e^{-2(br)^2} \rho_0^\alpha(br) \\ &\times \tilde{\rho}_{00,n'L'v'J',J'}^{J'0}(br)\, \tilde{\rho}_{mI,nLvJ,J'}^{J'0}(br)\,, \quad (121)\end{aligned}$$

where the density $\rho_0(br)$ depends on the reduced density as:

$$\rho_0(br) = \tilde{\rho}_{00,0000,00}^{00}(br)\frac{e^{-(br)^2}}{\sqrt{4\pi}}. \quad (122)$$

*4.7. Direct Coulomb energy in the spherical symmetry*

The integral for the direct Coulomb energy in spherical symmetry has a integrand which has a discontinuous first derivative. In order to get an integral which is easier to calculate using Gauss-Hermite integration we use the 'Vautherin trick' [8]

$$
\begin{aligned}
& \frac{e^2}{2}\int \mathrm{d}^3\boldsymbol{r}\,\mathrm{d}^3\boldsymbol{r'}\,\rho\left(r\right)\frac{1}{|\boldsymbol{r}-\boldsymbol{r'}|}\rho\left(r'\right) \\
= & \frac{\left(4\pi\right)^2 e^2}{6}\int \mathrm{d}\,r\,\mathrm{d}\,r'\,\left[\left(r+r'\right)^3-\left|r-r'\right|^3\right] rr'\rho\left(r\right)\Delta'\rho\left(r'\right) \\
= & \frac{4\pi e^2}{2\sqrt{3}}\int \mathrm{d}\,r\,\mathrm{d}\,r'\,\tilde{\rho}_{00,0000,0}\left(r\right) re^{-(br)^2} \\
& \times\int \mathrm{d}\,r\,\left[\left(r+r'\right)^3-\left|r-r'\right|^3\right] r'\tilde{\rho}_{20,0000,0}\left(r'\right) e^{-(br')^2} \\
= & \frac{4\pi e^2}{2\sqrt{3}}\int \tilde{\rho}_{00,0000,0}\left(r\right) re^{-(br)^2} V_{DC}\left(r\right)
\end{aligned}
\tag{123}
$$

which gives a smoother integrand. However in order to perform the linear-response calculations, where expressions for non-spherical multipoles are needed, we also developed a different method which will be presented elsewhere [9].

*4.8. Exchange Coulomb energy*

For the exchange Coulomb energy we use the Slater approximation

$$
\begin{aligned}
E_{ex} &= -\frac{3}{4}\left(\frac{3}{\pi}\right)^{1/3} e^2\int d\boldsymbol{r}\rho_p^{4/3} \\
&= -\frac{3}{4}\left(\frac{3}{\pi}\right)^{1/3} e^2 4\pi\int r^2 dr\tilde{\rho}_{p00,0000,0}\left(br\right)^{4/3}\frac{e^{-\frac{4}{3}(br)^2}}{\left(4\pi\right)^{2/3}} \\
&= -\frac{3^{4/3}e^2}{\left(4\right)^{2/3}}\int\left[\tilde{\rho}_{p00,0000,0}\left(br\right)^{4/3} e^{\frac{2}{3}(br)^2}\right] e^{-2(br)^2} r^2 dr.
\end{aligned}
$$

The matrix elements are the same as with density-dependent forces (using $\alpha = -2/3$ and $C^{\alpha,0000}_{00,0000} = -\frac{3}{4}\left(\frac{3}{\pi}\right)^{1/3} e^2$) but should only be added to the proton matrix elements.

*4.9. Numerical integration*

Numerical Gauss-Hermite integration is used to calculate the radial integrals occurring in the expressions for matrix elements [Eqs. (113) and (114)] and total energies [Eq. (120)]. This kind of integration is for integrals of the form $\int_{-\infty}^{\infty} e^{-x^2} f(x)\, dx$, and in order to obtain this form our integrals are transformed by using $x' = \left(\sqrt{2}b\right) x$. The integrals can then be written as:

$$\begin{aligned}\int_0^{\infty} e^{-2(bx)^2} f(bx)\, x^2 dx &= \frac{1}{2}\int_{-\infty}^{\infty} e^{-(x')^2} f_{sym}\left(\frac{x'}{\sqrt{2}}\right) \frac{x'^2}{2b^2}\frac{dx'}{\sqrt{2}b} \\ &= \frac{1}{2}\frac{1}{\sqrt{2}b^3}\int_{-\infty}^{\infty} e^{-(x')^2} f_{sym}\left(\frac{x'}{\sqrt{2}}\right)\left(\frac{x'}{\sqrt{2}}\right)^2 dx' \\ &\approx \frac{1}{\sqrt{2}b^3}\sum_{i=1}^{N/2} w_i f\left(x_i''\right)\left(x_i''\right)^2 .\end{aligned}$$

Where $f_{sym}(x) = f(x)\,\theta(x) + f(-x)\,\theta(-x)$ was used in an intermediate step. To reduce the number of grid points by half to $N_{\mathrm{grid}} = N/2$ it was also used that for Gauss-Hermite integration, the weight functions $w_i$ and grid points $x_i'' = \frac{x_i'}{\sqrt{2}} = br_i$ are symmetric about the origin. The integrals for matrix elements and total energies of most terms become exact when $N_{\mathrm{grid}} = N_0 + 2$, where $N_0$ denotes the maximum HO shell included in the basis. But in general more points are needed when the integrand cannot be expressed as a product of four basis states, e.g., in the case for the Coulomb interaction and also for the density-dependent terms.

## 5. Overview of the software structure

In order to get an idea of the software structure we list some of the main calls performed by the program. In the code HOSPHE (v1.00), there are also deeper level calls that are not listed here. Depending on the input data, some of the calls listed below may or may not be performed, but all are anyway included in the list. We specify the subroutine calls by first writing the name of the module followed by the name of the subroutine.

- hfmain

  Driver program which reads input and starts the run by calling the hf routine.

– hf

  Main program which contains the iteration loop. Starts by calling routines which initializes different tables.

  * RecoupDK: define_cfk
  * RecoupDK: define_cfd
  * RecoupA: define_cfa
  * RecoupA: define_cfa2
  * Coupling: restrict_densities
  * Coupling: get_cc_restrict
  * Coupling: make_ltab
  * Skyrme: get_skyrme_cc
  * Coupling: read_carlsson_from_file
  * Coupling: cc_perl_to_n3lo
  * Coupling: set_dependent_cc
  * Coupling: cc_n3lo_to_perl
  * grid: set_grid
  * qnotra: set_sps
  * qnotra: qpbase
  * mflib: Read_hf_initial_density_matrix
  * mflib: set_onedet
  * Ho_Derivatives: store_derivatives_of_basis
  * Ho_Derivatives: read_derivatives_of_basis
  * TypeDefinitions: inverse_derivative_densities
  * F2U: Define_F2U
  * mflib: read_density_matrix

  Next follows the main iteration loop.

  * Derivative_densities: sum_derivative_densities
  * hmatrix_Ud: hamiltonian_matrix
  * mflib: diagonalize_hf
  * mflib: hf2ho
  * density_energy: energy_from_densities
  * density_energy: energy_from_densities_densdep

* density_energy: energy_from_coulomb
* mflib: hf_energy

Finally there is some post-processing of the results.

* density_energy: plot_energy_terms2
* sorting: sortrx
* mflib: store_density_matrix

## 6. Description of the individual software components

The code HOSPHE (v1.00) is separated into different files with each file containing a module. Modules may contain module parameters and collections of subroutines and functions. Below follows a list of the files defining the code where the *'s should be replaced by the latest version number.

The modules which are used mainly in the initialization phase are:

- **hfmain_dist.v*.f90**

  Driver routine which reads the input and calls the hf subroutine.

- **hfmain_Parameters.v*.f90**

  Allocates parameters used as input to hf.

- **qnotra.v*.o**

  Defines quantum numbers for basis states and quantum numbers for the reduced density matrix.

- **skyrme.v*.f90**

  Contains coupling-constant sets for various standard Skyrme EDF's

- **coupling.v*.f90**

  Assigns quantum numbers to the coupling constants.

- **TypeDefinitions.v*.f90**

  Contains list of the allowed quantum numbers for the derivative operators, local densities, and derivative densities.

- **RecoupKlist.v*.f90**

  Contains one subroutine which reorders the list of $K$ coefficients.

- **RecoupDK.v*.f90**

  Contains values and quantum numbers of the $D$ and $K$ coefficients.

- **RecoupA.v*.f90**

  Contains values and quantum numbers of the $A$ coefficients which are used to recouple derivative densities.

- **F2U.v*.f90**

  Contains the $F2U$ coefficients. These are used to obtain fields for each term in the functional and are almost the same (see Table 7) as the $\chi$ coefficients in Eq. (8).

- **Ho_Derivatives.v*.f90**

  Contains routines to calculate derivatives of the HO basis functions.

- **grid.v*.f90**

  Generates Gauss-Hermite points and weights.

The modules which are used mainly in the iteration loop are:

- **hf.v*.f90**

  The main subroutine containing the iteration loop.

- **Derivative_densities.v*.f90**

  Generates the reduced derivative densities as functions of the grid points given a reduced density matrix as input.

- **hmatrix_Ud.v*.f90**

  Determines the matrix elements by using for the fields the form in Eq. (7). Calculates the reduced matrix elements given reduced densities and reduced matrix element quantum numbers as input. Also calculates the matrix elements of the density-dependent terms and of the Coulomb interaction.

- **matrix.v*.f90**

  Determines the matrix elements by using for the fields the form in Eq. (60). Calculates reduced matrix elements given reduced densities and reduced matrix element quantum numbers as input. Used mainly to test the accuracy of the matrix elements from hmatrix_Ud.v*.f90.

- **density_energy.v*.f90**

  Calculates integrals by using the Gauss-Hermite integration. Contains subroutines to calculate integrals over the terms in the functional to obtain the energy. Also calculates Coulomb energy integrals and radii.

- **coulomb.v*.f90**

  Calculates Coulomb direct and exchange matrix elements. Used mainly to test the accuracy of the matrix elements from hmatrix_Ud.v*.f90.

- **densitydep.v*.f90**

  Contains density-dependent factors that multiply each density-dependent coupling constants in the mean field potentials.

There are also modules containing helper routines:

- **sorting.v*.f90**

  Contains routines used to sort data.

- **geometric.v*.f90**

  Contains a collection of routines for the Clebsh-Gordon coefficients, 3, 6, and $9j$ symbols, Gamma functions etc.

- **mflib.v*.f90**

  Contains a collection of auxiliary routines such as the diagonalization routines and various transformations.

## 7. Differences between the notation in the code HOSPHE (v1.00) and in the text

In some cases, the notation used in this study differs from that used in the code HOSPHE (v1.00). For example, to store real objects in the code, we have removed complex phases from some coefficients. The transformation rules for some useful objects are defined in Table 7.

| Notation in text | Definition in the code | Stored in module |
|---|---|---|
| $\chi_{a,\alpha;\gamma}^{\beta;d\delta}$ | $(i)^{n_\gamma}\,(-i)^{n_a+n_d}$ F2U%Coeff | F2U_table |
| $F_{nLNl}^{l''}$ | $(-i)^n$F_val | Ho_Derivatives |
| $D_{mIm'I'}^{n'L'}$ | D_REC%Coeff | RecoupDK |
| $K_{mIm'I'}^{n'L'}$ | $(2i)^{-n'}\,(-1)^{-m'}$K_REC%Coeff | RecoupDK |
| $A_{mI,nLvJ,J'}^{uUu'U',W}$ | $(2i)^{-n}$A_REC%Coeff | RecoupA |
| $U_\gamma$ | $\frac{1}{2}\,(i)^{n_\gamma}$ U_fkn | hmatrix_Ud |

Table 1: Table relating notation introduced in the text with corresponding objects defined in the code HOSPHE (v1.00).

## 8. Description of input data

Input is given to the code HOSPHE (v1.00) by using the FORTRAN "namelist" statement. In this way, the variables specified in the input have their values assigned, while those not specified in the input retain their predefined default values. The variables that can be specified in the input are listed below.

1. **noscmax**
   Maximum main HO quantum number $N_0$ used in the HO basis. The code HOSPHE (v1.00) currently supports values of $N_0$ up to 70.

2. **ordermax**
   Maximum order in derivatives used. The possible choices are 0,2,4 and 6.

3. **ngrid**
   Number of Gauss-Hermite grid points. The code HOSPHE (v1.00) currently supports up to 85 grid points. If a negative value is given, the code uses $N_{\rm grid} = N_0 + 2 + 10$ points, that is, 10 more points than is needed for the most terms to become exact. However, for the Coulomb and density-dependent terms to converge with high precision, one may need more grid points.

4. **intera**
   Name of the Skyrme functional to be used. At present, supported versions are: "SLY4", "SLY5", "SKM*", "SKP", "SIII", and "FILE". If the name "FILE" is specified, the coupling constants in the spherical notation are read from file "cc.inp". An example file of this type is included in the distribution.

5. **AZ, AN**
   Number of protons and neutrons.

6. **hbarom**
   Oscillator frequency $\hbar\omega$ in MeV. If a negative value is given, it is calculated as $\hbar\omega = 1.2 \times 41A^{-1/3}$, where $A$ is the number of nucleons (see above).

7. **boscil**
   Oscillator constant $b$ in fm$^{-1}$. If a negative value is given, it is calculated from Eq. (80) by using the value of $\hbar\omega$ (see above) and the nucleon mass $m$ being the average of the neutron and proton masses.

8. **icm**
   Center of mass correction. For icm=0, no correction is used, and for icm=1, the code HOSPHE (v1.00) uses the one-body center of mass correction ($E_{\mathrm{c.m.}} \simeq -\frac{1}{A}\langle T\rangle$).

9. **icoudir**
   For icoudir=0, no direct Coulomb term is included, and for icoudir=$-1$, the code HOSPHE (v1.00) calculates the direct Coulomb energy by using the Vautherin method, see Section 4.7.

10. **icouex**
    For icouex=0, the Coulomb exchange term is not included, and for icouex=$-1$, the code HOSPHE (v1.00) calculates the Coulomb exchange energy by using the Slater approximation, see Section 4.8.

11. **itermax**
    Maximum number of iterations allowed before aborting.

12. **epsilon**
    Accuracy parameter. The iterations are stopped when the ground-state energies calculated by using the EDF and HF expressions differ by less

than epsilon and every HF single-particle energy changes less than epsilon between two iterations.

13. **alpha**
Mixing parameter $\alpha$ to slow-down/accelerate the iteration convergence. It mixes the density matrix from the current $\rho_1$ and previous $\rho_0$ iterations, so that the new density matrix is obtained as $\rho = \alpha * \rho_1 + (1 - \alpha) * \rho_0$

14. **keta_J**
Turns the Skyrme tensor coupling constants ON/OFF (1/0).

15. **restart**
If 1 or 2, the code HOSPHE (v1.00) attempts to read the density matrix from the file named as in the following example:
densities_050_082.rec for (AZ,AN)=(50,82) and the density matrix is also automatically stored to the same file when the iterations are finished. For restart = 0 no restart of iterations are attempted. For restart=2 the stored density matrix is used even though it may come from a calculation with a different number of oscillator shells. For restart=1 it is used only if the number of oscillator shells is the same.

16. **Flag_read_ini_dm**
If .true., the code HOSPHE (v1.00) attempts to read the initially occupied levels from the file "occ_orbs.inp". The first line should say $N_{\rm cng}, N_{\rm fill}$ where all main oscillator shells up to $N_{\rm fill}$ are automatically filled and $N_{\rm cng}$ denotes the number of changes in occupation with respect to this initial filling. Then follows one line per change, each change being specified as $N, l, 2j, I_{\rm occ}$ where $N, l, 2j$ denotes quantum numbers of HO $j$-shells and $I_{\rm occ}$ specifies weather the shell should be occupied ($I_{\rm occ} = 1$) or not ($I_{\rm occ} = 0$). This is repeated twice to define shell occupancies for protons and then for neutrons. An example file of this type is included in the distribution.

17. **verbose**
Verbose is an integer which specifies the amount of output produced by the code HOSPHE (v1.00) during a run. Verbose = 0 is the standard which gives a minimum of output and higher values leads to more information being printed to the screen.

## 9. Installation instructions

Most of the source code of HOSPHE (v1.00) is written in Fortran 95. The code consists of several modules that are linked together by using the standard Makefile (see [10]). In general, no special compiler options are needed, except for the optimization flags or check flags for tests. These options are set in the accompanying Makefile, which is now set to work with the G95 compiler that is a free and open-source Fortran 95 compiler (see [11]).

The code HOSPHE (v1.00) also make use of the LAPACK and BLAS linear-algebra libraries that can be obtained from [12, 13].

## 10. Test run description

The simplest way to run the code HOSPHE (v1.00) consists in providing the namelist input data in the following form:

```
./hosphe << end > pb208.n50.out
&input AN=126,AZ=82,noscmax=50,
icm=1,icoudir=-1,icouex=-1,keta_J=1,intera="SLy4",ordermax=2,
epsilon=1.e-7,itermax=1000,
Flag_read_ini_dm = .false.,restart = 0,
alpha=0.65,ngrid=-80,boscil=-2,hbarom=-1/
end
```

The script above is provided in the distribution file of the code HOSPHE (v1.00). By executing the script, one obtains the output file "pb208.n50.out", which is also provided in the distribution file. The main section of this file, which gives the total energies in $^{208}$Pb calculated for the maximum HO shell included in the basis of $N_0 = 50$ and SLy4 Skyrme functional [14], reads

```
    Kin.prot    Kin. neut.     Tot. kin.
 1337.059947   2529.116266   3866.176214

  T=0 Skyrme    T=1 Skyrme   Tot. Skyrme
-6405.081099    106.598348  -6298.482751
```

```
       Energy      HF Energy    Rearr. ene
 -1635.692396   -1635.692396  -1221.821085

    Cou. tot.      Cou. dir.     Cou. exc.
   796.614142     827.882912    -31.268770
```

In Figs. 1 and 2, we present results of similar calculations performed for $^{208}$Pb and the HO bases of $N_0$ = 10–70. Fig. 1 shows the convergence of the total energy in function of $N_0$. It turns out that the energy converges exponentially to the limiting value of $E_0$, namely,

$$E(N_0) = E_0 + E_1 \exp(-aN_0). \tag{124}$$

However, as shown in the two panels of Fig. 1, two different values of $E_0$ and $a$ are obtained for the regions of $N_0$ below and above $N_0 = 38$. A rather rapid convergence ($a = 0.172(3)$) to $E_0 = -1635.719$, which is seen below $N_0 = 38$, is followed by a slower convergence ($a = 0.1068(8)$) to $E_0 = -1635.69405$. Since the least-square fit of the limiting values $E_0$ is ill-conditioned, no error estimates can be obtained for them.

By considering convergence patterns in a few more cases for different options and nuclei, we found that the trend with two different slopes is not a general feature. In the few cases we looked at, we found that it is only above 40–50 shells that the rate seems to stabilize to an exponential convergence. These results show that, in general, its not possible to find the extrapolated limit of energy just by calculating only a few points of the curve for some small numbers of shells.

Fig. 2 shows the dependence of CPU times on $N_0$, obtained on the AMD Opteron Processor 2352 running at 2100 MHz clock speed. First, one can see that the spherical-basis code HOSPHE (v1.00) is, of course, orders of magnitude faster that the 3D code HFODD (v2.40h) [15]. For $N_0 = 36$, the former needs only 20 sec of CPU time while the latter needs 250 000 sec. Second, for both codes, the dependencies on $N_0$ are clearly given by power lows indicated in the figure. Strangely enough, these power lows are different for calculations performed below and above $N_0 = 20$. At the moment, no explanation for such a timing pattern could be found.

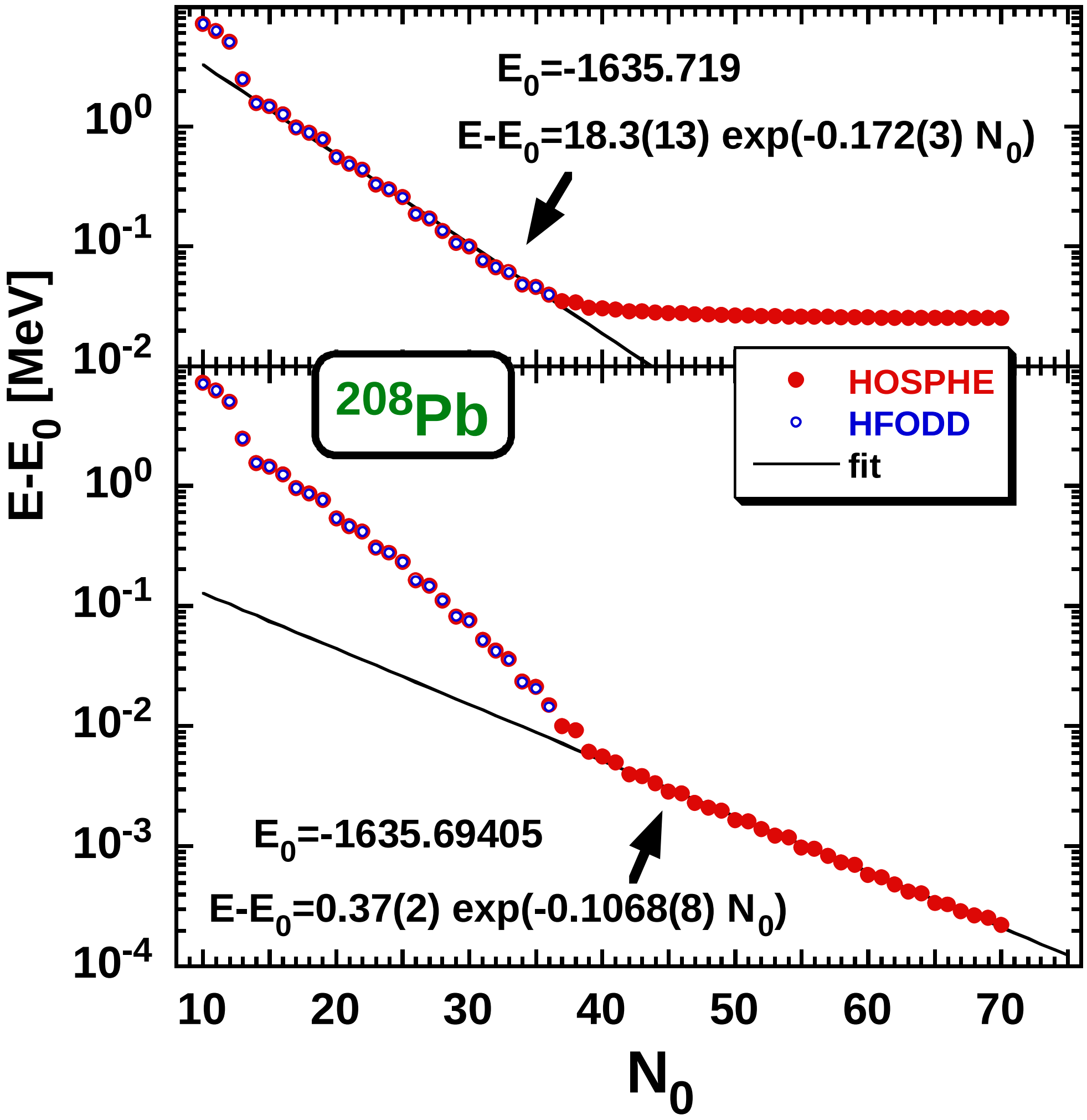

Figure 1: Total energy of $^{208}$Pb as a function of the maximum HO shell included in the basis $N_0$. Upper and lower panels show in the logarithmic scale differences $E(N_0) - E_0$ for two different values of $E_0$. Large full dots and small empty circles give results calculated by using the codes HOSPHE (v1.00) and HFODD (v2.40h) [15], respectively. Up to $N_0 = 36$, where the HFODD calculations could have been performed, one sees a perfect agreement between the results given by the two codes. Solid lines give results of the exponential-decay fits.

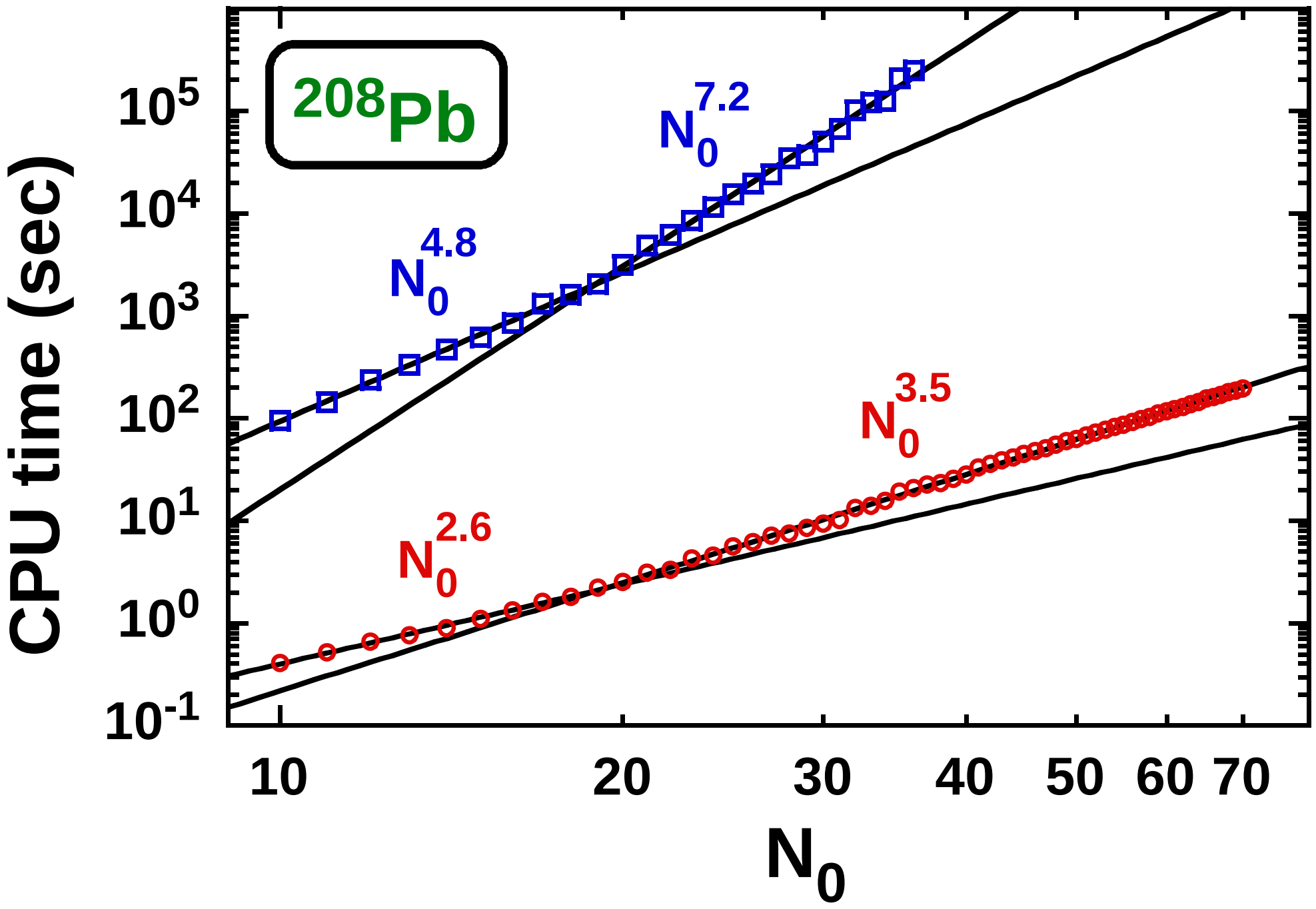

Figure 2: The HOSPHE (v1.00) CPU times required for calculations performed for $^{208}$Pb and the standard Skyrme functional SLy4, shown as functions of the maximum HO shell included in the basis $N_0$. Circles and squares show results obtained by using the codes HOSPHE (v1.00) and HFODD (v2.40h) [15], respectively. The doubly logarithmic scale in the Figure, shows that these times scale as different powers of $N_0$, as indicated in the figure.

## 11. Summary

We have developed a computer code that is able to solve the self-consistent equations resulting from the use of the generalized N$^3$LO energy density functional introduced in Ref. [4]. For different terms of the functional, the resulting mean-field expressions were systematically formulated by using the spherical-tensor representation, which was also introduced in Ref. [4]. In Section 3 we presented the general equations, needed for any choice of basis, and in Section 4 we developed expressions needed to implement the formalism in the spherical harmonic oscillator basis.

The present version of the program is based on the spherical harmonic oscillator basis and can be used to calculate, for example, ground state properties of spherical nuclei using standard Skyrme forces. For this purpose, it has some advantage compared to other codes, because of its fast execution time and stability.

The motivation to construct the code, and its main new feature, is the possibility to test the influence of higher order derivative terms included in the mean field. The values of the higher order coupling constants can be calculated from the density-matrix expansions starting from other interactions or by direct fits to experimental data, and both approaches are presently being explored. The formalism was derived in a general way, so as to include non-spherical multipoles, and thus it can also be used for linear-response calculations [16], which go beyond the mean field. Extensions to include pairing, better treatment of the continuum, and statically deformed densities are also being developed.

## 12. Acknowledgements

This work was supported by the Academy of Finland and University of Jyväskylä within the FIDIPRO program and by the Polish Ministry of Science and Higher Education under Contract No. N N 202 328234.

## A. Calculation of coefficients $D^{uU}_{rRpP}$ (72)

Coefficients $D^{uU}_{rRpP}$ (72) can easily be calculated by using the fact that they exactly correspond to the multiplication of tensors $X$ built from the position vector $x$.

$$X_{rRM_R}X_{pPM_P} = \sum_{uUM_U} C^{UM_U}_{RM_RPM_P} D^{uU}_{rRpP} X_{uUM_U}. \tag{125}$$

Tensors $X$ are simply related to spherical harmonics (see Eq. 5.2(40) in Ref. [6]):

$$X_{pPM_P} = \left(-\frac{r^2}{\sqrt{3}}\right)^{\frac{p-P}{2}} r^P \sqrt{\frac{4\pi P!}{(2P+1)!!}} Y_{PM_P}(\theta,\phi). \tag{126}$$

Therefore, multiplication law of spherical harmonics (Eq. 5.6(9) in Ref. [6]) gives

$$\begin{aligned} X_{rRM_R}X_{pPM_P} &= \left(-\sqrt{3}\right)^{\frac{R-r}{2}} r^r \sqrt{\frac{4\pi R!}{(2R+1)!!}} \\ &\times \left(-\sqrt{3}\right)^{\frac{P-p}{2}} r^p \sqrt{\frac{4\pi P!}{(2P+1)!!}} \sum_{UM_U} \sqrt{\frac{(2R+1)(2P+1)}{4\pi(2U+1)}} \\ &\times C^{U0}_{R0P0} C^{UM_U}_{RM_RPM_P} \left(-\sqrt{3}\right)^{\frac{u-U}{2}} r^{-u} \sqrt{\frac{(2U+1)!!}{4\pi U!}} X_{uUM_U}, \end{aligned} \tag{127}$$

and thus the required coefficients read

$$D^{uU}_{rRpP} = \delta_{u,r+p} \left(-\sqrt{3}\right)^{\frac{R+P-U}{2}} \sqrt{\frac{R!P!}{U!}\frac{(2U-1)!!}{(2R-1)!!(2P-1)!!}} C^{U0}_{R0P0}, \tag{128}$$

and are independent of $urp$.

## B. Derivation of Eq. (98)

We begin by a summation of the product of four Clebsh-Gordan coefficients, that appear in Eq. (97), that is,

$$\begin{aligned} &\sum_{mm_sm'm'_s} C^{\frac{1}{2}m'_s}_{\frac{1}{2}m_sv\mu} C^{jm}_{lm_l\frac{1}{2}m_s} C^{jm}_{j'm'J''M''} C^{j'm'}_{l'm'_l\frac{1}{2}m'_s} \\ = &\sum_{mm_sm'm'_s} (-1)^{\frac{1}{2}-m_s} \sqrt{\frac{2}{2v+1}} C^{v,-\mu}_{\frac{1}{2}m_s\frac{1}{2},-m'_s} (-1)^{\frac{1}{2}+m_s} \sqrt{\frac{2j+1}{2l+1}} C^{l,-m_l}_{j,-m\frac{1}{2}m_s} \\ &\times(-1)^{j'-m'} \sqrt{\frac{2j+1}{2J''+1}} C^{J'',-M''}_{j'm'j,-m} (-1)^{\frac{1}{2}+m'_s} \sqrt{\frac{2j'+1}{2l'+1}} C^{l'm'_l}_{\frac{1}{2},-m'_sj'm'} \\ = &(-1)^{\frac{1}{2}}(-1)^{\frac{1}{2}}(-1)^{j'-m'_l}(-1)^{\frac{1}{2}} \sqrt{\frac{2}{2v+1}} \sqrt{\frac{2j+1}{2l+1}} \sqrt{\frac{2j+1}{2J''+1}} \sqrt{\frac{2j'+1}{2l'+1}} \\ &\times(-1)^{1-v}(-1)^{j+\frac{1}{2}-l} \sum_{mm_sm'm'_s} C^{J'',-M''}_{j'm'jm} C^{v,-\mu}_{\frac{1}{2}m'_s\frac{1}{2}m_s} C^{l'm'_l}_{\frac{1}{2}m'_sj'm'} C^{l,-m_l}_{\frac{1}{2}m_sjm} \end{aligned}$$

$$
\begin{aligned}
&= (-1)^{j'-m_l'+j-l+1-v}\sqrt{2}\sqrt{2j+1}\sqrt{2j+1}\sqrt{2j'+1} \\
&\times \sum_{T'M_T'} C^{T'M_T'}_{l'm_l'l,-m_l} C^{T'M_T'}_{v,-\mu J'',-M''} \left\{ \begin{array}{ccc} j & j' & J'' \\ \frac{1}{2} & \frac{1}{2} & v \\ l & l' & T' \end{array} \right\}, \qquad (129)
\end{aligned}
$$

see Eq. 8.7(20) in Ref. [6].

The product of spherical harmonics reads

$$
\begin{aligned}
Y_{km_k}(\theta,\phi)Y^*_{k'm_k'}(\theta,\phi) &= (-1)^{m_k'} \sum_{TM_T} \sqrt{\tfrac{(2k+1)(2k'+1)}{4\pi(2T+1)}} \\
&\times C^{T0}_{k0k'0} C^{TM_T}_{km_kk',-m_k'} Y_{TM_T}(\theta,\phi), \qquad (130)
\end{aligned}
$$

see Eqs. 5.4(1) and 5.6(9) in Ref. [6]. This gives us another sum of products of four Clebsh-Gordan coefficients to sum up:

$$
\begin{aligned}
&\sum_{m_l m_l' m_k m_k'} (-1)^{m_k'-m_l'} C^{km_k}_{lm_lUM_U} C^{k'm_k'}_{l'm_l'U',-M_U'} C^{T'M_T'}_{l'm_l'l,-m_l} C^{TM_T}_{km_kk',-m_k'} \\
&= \sum_{m_l m_l' m_k m_k'} (-1)^{m_k'-m_l'} (-1)^{l-m_l} \sqrt{\tfrac{2k+1}{2U+1}} C^{UM_U}_{km_kl,-m_l} \\
&\times (-1)^{l'-m_l'} \sqrt{\tfrac{2k'+1}{2U'+1}} C^{U'M_U'}_{l'm_l'k',-m_k'} C^{T'M_T'}_{l'm_l'l,-m_l} C^{TM_T}_{km_kk',-m_k'} \\
&= (-1)^{M_T'-M_U'} (-1)^{l} (-1)^{l'} \sqrt{\tfrac{2k+1}{2U+1}} \sqrt{\tfrac{2k'+1}{2U'+1}} (-1)^{l'+k'-U'} (-1)^{k+k'-T} \\
&\times \sum_{m_l m_l' m_k m_k'} C^{UM_U}_{km_kl,-m_l} C^{U'M_U'}_{k',-m_k'l'm_l'} C^{T'M_T'}_{l'm_l'l,-m_l} C^{TM_T}_{k',-m_k'km_k} \\
&= (-1)^{M_T'-M_U'} (-1)^{l+k-U'-T} \sqrt{\tfrac{2k+1}{2U+1}} \sqrt{\tfrac{2k'+1}{2U'+1}} \\
&\times \sum_{m_l m_l' m_k m_k'} C^{UM_U}_{km_klm_l} C^{U'M_U'}_{k'm_k'l'm_l'} C^{TM_T}_{k'm_k'km_k} C^{T'M_T'}_{l'm_l'lm_l} \\
&= (-1)^{M_T'-M_U'} (-1)^{l+k-U'-T} \sqrt{(2k+1)(2k'+1)(2T'+1)(2T+1)} \\
&\times \sum_{W'M_W'} C^{W'M_W'}_{TM_TT'M_T'} C^{W'M_W'}_{U'M_U'UM_U} \left\{ \begin{array}{ccc} l & k & U \\ l' & k' & U' \\ T' & T & W' \end{array} \right\}, \qquad (131)
\end{aligned}
$$

see Eq. 8.7(20) in Ref. [6].

We can now perform the summation over $M_U'$ and $M_U$, which gives the factor $(-1)^{U+U'-W}\delta_{WW'}\delta_{M_WM_W'}$ and allows for a summation over $W'$ and

$M'_W$. After inserting all these results into Eq. (97), we obtain

$$
\begin{aligned}
\tilde{\rho}^{uUu'U'W,J''M''}_{vJ'M'}(\boldsymbol{r}) = \; & \sum_{M_W\mu} C^{J'M'}_{WM_Wv\mu}(-1)^{U'} \sum_{\substack{Nlj\\N'l'j'}} b^{3+u+u'} e^{-(br)^2} \tfrac{1}{\sqrt{2}} \langle \tfrac{1}{2}||\sigma_v||\tfrac{1}{2}\rangle \\
\times \; & \sum_{k} \tfrac{1}{\sqrt{2k+1}} F^{k}_{uUNl}(br) \tfrac{1}{\sqrt{2j+1}} \langle \phi_{Nlj}||\tilde{\rho}^{J''}||\phi_{N'l'j'}\rangle \\
\times \; & \sum_{k'} \tfrac{1}{\sqrt{2k'+1}} F^{k'*}_{u'U'N'l'}(br) \\
\times \; & (-1)^{j'+j-l+1-v} \sqrt{2}\sqrt{2j+1}\sqrt{2j+1}\sqrt{2j'+1} \\
& \times \sum_{T'M'_T} C^{T'M'_T}_{v,-\mu J'',-M''} \begin{Bmatrix} j & j' & J'' \\ \frac{1}{2} & \frac{1}{2} & v \\ l & l' & T' \end{Bmatrix} \\
\times \; & \sum_{TM_T} \sqrt{\tfrac{(2k+1)(2k'+1)}{4\pi(2T+1)}} \times C^{T0}_{k0k'0} Y_{TM_T}(\theta,\phi) \\
\times \; & (-1)^{M'_T+l+k-U'-T} \sqrt{(2k+1)(2k'+1)(2T'+1)(2T+1)} \\
& \times C^{WM_W}_{TM_TT'M'_T} \begin{Bmatrix} l & k & U \\ l' & k' & U' \\ T' & T & W \end{Bmatrix} (-1)^{U+U'-W}. \qquad (132)
\end{aligned}
$$

Now we have to sum up products of three Clebsh-Gordan coefficients:

$$
\begin{aligned}
& \sum_{M_WM'_T\mu} (-1)^{M'_T} C^{J'M'}_{WM_Wv\mu} C^{T'M'_T}_{v,-\mu J'',-M''} C^{WM_W}_{TM_TT'M'_T} \\
= \; & \sum_{M_WM'_T\mu} (-1)^{M'_T} C^{J'M'}_{WM_Wv\mu} (-1)^{v+\mu} \sqrt{\tfrac{2T'+1}{2J''+1}} C^{J''M''}_{v,-\mu T',-M'_T} \\
& \times (-1)^{T'+M'_T} \sqrt{\tfrac{2W+1}{2T+1}} C^{TM_T}_{WM_WT',-M'_T} (-1)^{W+T'-T} \\
= \; & \sqrt{\tfrac{2T'+1}{2J''+1}} (-1)^{T'} \sqrt{\tfrac{2W+1}{2T+1}} (-1)^{W+T'-T} \\
& \times \sum_{M_WM'_T\mu} (-1)^{v-\mu} C^{J''M''}_{v\mu T'M'_T} C^{TM_T}_{WM_WT'M'_T} C^{J'M'}_{WM_Wv,-\mu} \\
= \; & \sqrt{\tfrac{2T'+1}{2J''+1}} (-1)^{T'} \sqrt{\tfrac{2W+1}{2T+1}} (-1)^{W+T'-T} \\
& \times (-1)^{T'+W+J''+J'} \sqrt{2J''+1}\sqrt{2J'+1} C^{TM_T}_{J''M''J'M'} \begin{Bmatrix} v & T' & J'' \\ T & J' & W \end{Bmatrix}, \qquad (133)
\end{aligned}
$$

see Eq. 8.7(15) in Ref. [6]. This gives:

$$\begin{aligned}
\tilde{\rho}^{uUu'U'W,J''M''}_{vJ'M'}(\boldsymbol{r}) &= (-1)^{U'} \sum_{\substack{Nlj\\N'l'j'}} b^{3+u+u'} e^{-(br)^2} \tfrac{1}{\sqrt{2}} \langle \tfrac{1}{2} || \sigma_v || \tfrac{1}{2} \rangle \\
&\times \sum_k \tfrac{1}{\sqrt{2k+1}} F^k_{uUNl}(br) \tfrac{1}{\sqrt{2j+1}} \langle \phi_{Nlj} || \tilde{\rho}^{J''} || \phi_{N'l'j'} \rangle \\
&\times \sum_{k'} \tfrac{1}{\sqrt{2k'+1}} F^{k'*}_{u'U'N'l'}(br) \\
&\times (-1)^{j'+j-l+1-v} \sqrt{2}\sqrt{2j+1}\sqrt{2j+1}\sqrt{2j'+1} \\
&\times \sum_{T'} \left\{ \begin{array}{ccc} j & j' & J'' \\ \frac{1}{2} & \frac{1}{2} & v \\ l & l' & T' \end{array} \right\} \\
&\times \sum_{TM_T} \sqrt{\tfrac{(2k+1)(2k'+1)}{4\pi(2T+1)}} \times C^{T0}_{k0k'0} Y_{TM_T}(\theta,\phi) \\
&\times (-1)^{l+k-U'-T} \sqrt{(2k+1)(2k'+1)(2T'+1)(2T+1)} \\
&\times \left\{ \begin{array}{ccc} l & k & U \\ l' & k' & U' \\ T' & T & W \end{array} \right\} (-1)^{U+U'-W} \\
&\times \sqrt{\tfrac{2T'+1}{2J''+1}} (-1)^{T'} \sqrt{\tfrac{2W+1}{2T+1}} (-1)^{W+T'-T} (-1)^{T'+W+J''+J'} \\
&\times \sqrt{2J''+1}\sqrt{2J'+1} C^{TM_T}_{J''M''J'M'} \left\{ \begin{array}{ccc} v & T' & J'' \\ T & J' & W \end{array} \right\}
\end{aligned} \quad (134)$$

Finally we have:

$$\begin{aligned}
\tilde{\rho}^{uUu'U'W,J''M''}_{vJ'M'}(\boldsymbol{r}) &= (-1)^{1+v} b^{3+u+u'} e^{-(br)^2} \langle \tfrac{1}{2} || \sigma_v || \tfrac{1}{2} \rangle \sqrt{\tfrac{1}{4\pi}} \\
&\times \sum_{\substack{Nljk\\N'l'j'k'}} F^k_{uUNl}(br) \langle \phi_{Nlj} || \tilde{\rho}^{J''} || \phi_{N'l'j'} \rangle F^{k'}_{u'U'N'l'}(br) \\
&\times \sum_{T'TM_T} (-1)^{j'+j} (-1)^k (-1)^{T+T'} (-1)^{U-W} C^{T0}_{k0k'0} \\
&\times \sqrt{\tfrac{(2T'+1)(2T'+1)(2k+1)(2k'+1)(2J'+1)(2W+1)(2j+1)(2j'+1)}{(2T+1)}} \\
&\times \left\{ \begin{array}{ccc} j & j' & J'' \\ \frac{1}{2} & \frac{1}{2} & v \\ l & l' & T' \end{array} \right\} \left\{ \begin{array}{ccc} l & k & U \\ l' & k' & U' \\ T' & T & W \end{array} \right\} \left\{ \begin{array}{ccc} v & T' & J'' \\ T & J' & W \end{array} \right\}
\end{aligned}$$

$$\times \quad C^{TM_T}_{J'M'J''M''} Y_{TM_T}(\theta,\phi), \tag{135}$$

where we have also used Eq. (90). This gives expression (98) and definitions of radial form factors (99) and coefficients (100).

### C. Derivation of Eq. (113)

After inserting Eq. (111) into Eq. (112) we must sum up products of four Clebsh-Gordan coefficients, similarly as in Eq. (129), that is,

$$\begin{aligned}
&\sum_{m_j m_s m'_j m'_s} C^{\frac{1}{2}m'_s}_{\frac{1}{2}m_s v'\mu'} C^{jm_j}_{lm_l\frac{1}{2}m_s} C^{j'm'_j}_{jm_jI''M''_I} C^{j'm'_j}_{l'm'_l\frac{1}{2}m'_s} \\
= \;& (-1)^{M''_I - m'_l - l - v'} \sqrt{2}\sqrt{2j'+1}\sqrt{2j'+1}\sqrt{2j+1} \\
& \times \sum_{T'M'_T} C^{T'M'_T}_{l'm'_l l,-m_l} C^{T'M'_T}_{v',-\mu' I''M''_I} \left\{ \begin{array}{ccc} j & j' & I'' \\ \frac{1}{2} & \frac{1}{2} & v' \\ l & l' & T' \end{array} \right\},
\end{aligned} \tag{136}$$

and then as in Eq. (131):

$$\begin{aligned}
&\sum_{m_l m'_l m_k m'_k} (-1)^{-m'_l} C^{km_k}_{lm_lIM_I} C^{k'm'_k}_{l'm'_lI',-M'_I} C^{T'M'_T}_{l'm'_l l,-m_l} C^{k'm'_k}_{TM_Tkm_k} \\
= \;& (-1)^{l+k+k'-I'} \sqrt{(2k+1)(2k'+1)(2k'+1)(2T'+1)} \\
& \times \sum_{W'M'_W} (-1)^{M'_W - M'_I} C^{W'M'_W}_{T,-M_TT'M'_T} C^{W'M'_W}_{I'M'_IIM_I} \left\{ \begin{array}{ccc} l & k & I \\ l' & k' & I' \\ T' & T & W' \end{array} \right\}.
\end{aligned} \tag{137}$$

We can now perform the summation over $M'_I$ and $M_I$, which gives the factor $(-1)^{I+I'-L'}\delta_{L'W'}\delta_{M'_LM'_W}$ and allows for a summation over $W'$ and $M'_W$. After inserting all these results into Eq. (112), we obtain

$$\begin{aligned}
\langle\phi_{N'l'j'}||\tilde{h}^{I''}(\tilde{\rho}^{J''M''})||\phi_{Nlj}\rangle = & \sum_{M''_I} \sum_{n'L'v'J'} \sum_{mIm'I'} \sum_{M'_L\mu'} \int r^2 \mathrm{d}\, r \\
& \sum_{M'} C^{J'M'}_{L'M'_Lv'\mu'} \tfrac{(-1)^{J'-M'}}{\sqrt{2J'+1}} (-1)^{m'} K^{n'L'}_{mIm'I'} (-1)^{m'+I'} \\
\times \; & \sum_{TM_T} \sum_{k'} \sum_{k} \tfrac{1}{\sqrt{2}} \langle\tfrac{1}{2}||\sigma_{v'}||\tfrac{1}{2}\rangle \sqrt{\tfrac{(2T+1)(2k+1)}{4\pi}} C^{k'0}_{T0k0}
\end{aligned}$$

$$\begin{aligned}
&\times\quad b^{3+n'} F^{k'*}_{m'I'N'l'}(br)\tilde{U}^{TJ''}_{n'L'v'J'}(br)F^{k}_{mINl}(br)e^{-2(br)^2}\\
&\times\quad C^{TM_T}_{J',-M'J''M''}\\
&\times\quad (-1)^{M''_I-l-v'}\sqrt{2}\sqrt{2j'+1}\sqrt{2j+1}\\
&\quad\times\sum_{T'M'_T} C^{T'M'_T}_{v',-\mu' I''M''_I}\left\{\begin{array}{ccc} j & j' & I'' \\ \frac{1}{2} & \frac{1}{2} & v' \\ l & l' & T' \end{array}\right\}\\
&\times\quad (-1)^{M'_L}(-1)^{l+k+k'-I'}\sqrt{(2T'+1)}\\
&\quad\times C^{L'M'_L}_{T,-M_T T'M'_T}\left\{\begin{array}{ccc} l & k & I \\ l' & k' & I' \\ T' & T & L' \end{array}\right\}(-1)^{I+I'-L'}, \qquad (138)
\end{aligned}$$

Now, we have to sum up products of three Clebsh-Gordan coefficients, similarly as in Eq. (133), that is,

$$\begin{aligned}
&\sum_{M'_L M'_T \mu'}(-1)^{M'_L} C^{J'M'}_{L'M'_L v'\mu'} C^{T'M'_T}_{v',-\mu' I''M''_I} C^{L'M'_L}_{T,-M_T T'M'_T}\\
=\;& \sqrt{\tfrac{2T'+1}{2I''+1}}(-1)^{T'}\sqrt{\tfrac{2L'+1}{2T+1}}(-1)^{L'+T'-T+M_T}\\
&\times(-1)^{T'+L'+I''+J'}\sqrt{2I''+1}\sqrt{2J'+1}C^{T,-M_T}_{I'',-M''_I J'M'}\left\{\begin{array}{ccc} v' & T' & I'' \\ T & J' & L' \end{array}\right\} \qquad (139)
\end{aligned}$$

This gives:

$$\begin{aligned}
&\langle\phi_{N'l'j'}||\tilde{h}^{I''}(\tilde{\rho}^{J''M''})||\phi_{Nlj}\rangle\\
&\quad=\sum_{M''_I}\sum_{n'L'v'J'}\sum_{mIm'I'}\sum_{M'}(-1)^{J'}K^{n'L'}_{mIm'I'}(-1)^{I'}\\
&\quad\times\quad \sum_{TM_T}\sum_{k'}\sum_{k}\tfrac{1}{\sqrt{2}}\langle\tfrac{1}{2}||\sigma_{v'}||\tfrac{1}{2}\rangle\sqrt{\tfrac{(2k+1)}{4\pi}}C^{k'0}_{T0k0}\\
&\quad\times\quad \int r^2\mathrm{d}\,r\, b^{3+n'}F^{k'*}_{m'I'N'l'}(br)\tilde{U}^{TJ''}_{n'L'v'J'}(br)F^{k}_{mINl}(br)e^{-2(br)^2}\\
&\quad\times\quad C^{TM_T}_{J',-M'J''M''}C^{T,-M_T}_{I'',-M''_I J'M'}\\
&\quad\times\quad (-1)^{I''+J'}(-1)^{T'}(-1)^{T}(-1)^{k+k'-I'-v'}(-1)^{I+I'-L'}\\
&\quad\times\quad \sqrt{2}\sqrt{2j'+1}\sqrt{2j+1}\sqrt{(2T'+1)}\sqrt{(2L'+1)(2T'+1)}\\
&\quad\times\quad \sum_{T'}\left\{\begin{array}{ccc} j & j' & I'' \\ \frac{1}{2} & \frac{1}{2} & v' \\ l & l' & T' \end{array}\right\}\left\{\begin{array}{ccc} l & k & I \\ l' & k' & I' \\ T' & T & L' \end{array}\right\}\left\{\begin{array}{ccc} v' & T' & I'' \\ T & J' & L' \end{array}\right\}. \qquad (140)
\end{aligned}$$

Finally, summations over $M'$ and $M_T$ give the factor $\delta_{I''J''}\delta_{M_I''M''}$, that is:

$$
\begin{aligned}
&\langle\phi_{N'l'j'}||\tilde{h}^{I''}(\tilde{\rho}^{J''M''})||\phi_{Nlj}\rangle \\
&= \delta_{I''J''}(-1)^{J''}\sum_{n'L'v'J'}\sum_{mIm'I'}K^{n'L'}_{mIm'I'}(-1)^{v'}\langle\tfrac{1}{2}||\sigma_{v'}||\tfrac{1}{2}\rangle\sqrt{\tfrac{1}{4\pi}} \\
&\times\sum_{Tkk'}(-1)^{T}(2T+1)\int r^2\mathrm{d}\,r\,b^{3+n'}F^{k'}_{m'I'N'l'}(br)\tilde{U}^{TJ''}_{n'L'v'J'}(br)F^{k}_{mINl}(br)e^{-2(br)^2} \\
&\times(-1)^{k}(-1)^{T}(-1)^{I-L'}C^{T0}_{k0k'0}\sqrt{\tfrac{(2k+1)(2k'+1)(2L'+1)(2j+1)(2j'+1)}{(2T+1)}} \\
&\times\sum_{T'}(-1)^{T'}(2T'+1)\left\{\begin{array}{ccc} j & j' & J'' \\ \frac{1}{2} & \frac{1}{2} & v' \\ l & l' & T'\end{array}\right\}\left\{\begin{array}{ccc} l & k & I \\ l' & k' & I' \\ T' & T & L'\end{array}\right\}\left\{\begin{array}{ccc} v' & T' & J'' \\ T & J' & L'\end{array}\right\},
\end{aligned}
\tag{141}
$$

where we have also used Eq. (90). This shows explicitly that in the field calculated for the density matrix with multipolarity $J''$, only the multipole $J''$ appears.